\documentclass[%
 reprint,
groupedaddress,
 amsmath,amssymb,
 aps,
floatfix
]{revtex4-1}

\usepackage{graphicx}
\usepackage{dcolumn}
\usepackage{bm}

\usepackage{xcolor}
\usepackage{pgfplots}
\usepgfplotslibrary{fillbetween}
\usepackage{siunitx}

\usepackage{tikz}
\usetikzlibrary{positioning,shapes,calc}

\usepackage{diagbox}

\usepackage{braket}
\usepackage{dsfont}

\newcommand{\catA}{{\color{uobRed}$\blacksquare$}}
\newcommand{\catB}{{\color{uobBrightYellow}$\blacktriangle$}}
\newcommand{\catC}{\tikz{\draw[uobDarkGreen,fill=uobDarkGreen](0,0) circle (.5ex);}}

\definecolor{uobRed}{RGB}{192,47,56}
\definecolor{uobStone}{RGB}{190,183,169}
\definecolor{uobBrightAqua}{RGB}{0,192,181}
\definecolor{uobDarkAqua}{RGB}{0,61,76}
\definecolor{uobBrightBlue}{RGB}{12,198,222}
\definecolor{uobDarkBlue}{RGB}{0,47,95}
\definecolor{uobBrightPurple}{RGB}{146,120,209}
\definecolor{uobDarkPurple}{RGB}{66,20,95}
\definecolor{uobBrightPink}{RGB}{224,36,154}
\definecolor{uobDarkPink}{RGB}{119,32,89}
\definecolor{uobBrightRed}{RGB}{224,0,52}
\definecolor{uobDarkRed}{RGB}{94,48,50}
\definecolor{uobBrightYellow}{RGB}{255,182,18}
\definecolor{uobDarkYellow}{RGB}{134,67,30}
\definecolor{uobBrightLime}{RGB}{190,214,0}
\definecolor{uobDarkLime}{RGB}{83,104,43}
\definecolor{uobBrightGreen}{RGB}{52,178,51}
\definecolor{uobDarkGreen}{RGB}{2,71,49}
\definecolor{ccb1}{HTML}{003d4c}
\definecolor{ccb2}{HTML}{00c0b5}
\definecolor{ccb3}{HTML}{34b233}
\definecolor{ccb4}{HTML}{ffb612}
\definecolor{ccb5}{HTML}{e00034}

\DeclareMathOperator{\SNR}{SNR}
\DeclareMathOperator{\Trace}{Tr}

\begin{document}


\title{Quantum Rangefinding}

\author{Stefan Frick}
\email{stefan.frick@uibk.ac.at}
\affiliation{Institut für Experimentalphysik, Universität Innsbruck, Technikerstra{\ss}e 25, 6020
  Innsbruck, Austria}
\affiliation{Quantum Engineering Technology Labs, H. H. Wills Physics Laboratory and Department of Electrical and
  Electronic Engineering, University of Bristol, BS8 1FD, UK}
\affiliation{Quantum Engineering Centre for Doctoral Training, Nanoscience and Quantum
  Information Centre, University of Bristol, BS8 1FD, United Kingdom}

\author{Alex McMillan}%
\affiliation{Quantum Engineering Technology Labs, H. H. Wills Physics Laboratory and Department of Electrical and
  Electronic Engineering, University of Bristol, BS8 1FD, UK}

\author{John Rarity}
\affiliation{Quantum Engineering Technology Labs, H. H. Wills Physics Laboratory and Department of Electrical and
Electronic Engineering, University of Bristol, BS8 1FD, UK}

\date{\today}

\begin{abstract}
Quantum light generated in non-degenerate squeezers has many applications such as sub-shot-noise
transmission measurements to maximise the information extracted by one photon or quantum
illumination to increase the probability in target detection.
However, any application thus far fails to consider the thermal characteristics of one half of the
bipartite down-converted photon state often used in these experiments.
We show here that a maximally mixed state, normally viewed as nuisance, can indeed be used to
extract information about the position of an object while at the same time providing efficient
camouflaging against other thermal or background light.
\end{abstract}

\maketitle

\section{Introduction}\label{sec:introduction}

Successfully harnessing the properties of quantum states of light has become one of the greatest
promises of quantum optics to revolutionise
computation~\cite{O'Brien2007,O'Brien2003,O'Brien2009}, sensing~\cite{Pirandola2018},
imaging~\cite{Genovese2016} and metrology~\cite{Adesso2016}.
This promise has been realised for many applications in metrology~\cite{Moreau2017,Whittaker2017},
sensing~\cite{Lopaeva2013}, and partially for computation~\cite{Shadbolt2011}.
However, all applications to date fail to realise any quantum advantage when, as in every realistic
system, loss and background are introduced.

It was realised in the early 90's that the energy-time correlations of pair photon sources could be
used to establish optical communication~\cite{Seward1991} and rangefinding~\cite{Rarity1990} (LIDAR)
in high background situations.
These early works exploited the time correlations of photon pairs allowing heralding of sent
photons, thus suppressing uncorrelated background and outperforming weak coherent state pulses when
adjusted to equivalent levels of sent photons per pulse, an advantage recently quantified
in~\cite{Liu2019}.
However, the time correlations of photon pairs are purely classical and any scheme using them to
gain an advantage can be implemented just as efficiently with classical pulsed light sources.
Additionally, this approach is limited to a maximum of one photon per heralded time gate,
constraining these approaches to low loss ($<\, -50\, \si{\deci\bel}$) scenarios not typically
encountered.

This advantage was again highlighted by Lloyd in 2008, who extended this to a theoretical framework
where light entangled over $n$ modes is used to illuminate a target~\cite{Lloyd2008}.
This entangled quantum illumination promised suppression of false detection probabilities by a
factor of $n$ even if loss is present in the system.
Later, this prediction of quantum illumination was softened when comparing the performance of
entangled photons to coherent~\cite{Shapiro2009}, non-Gaussian~\cite{Zhang2014} and
Gaussian~\cite{Tan2008} states.
With the latter work proving that quantum illumination can never achieve an advantage bigger than
$6\,\si{\deci\bel}$ \cite{Shapiro2020}.
While detectors achieving $3\,\si{\deci\bel}$ advantage can be implemented with relative
ease~\cite{Guha2009,Zheshen2015} a detector achieving the full advantage of quantum illumination is
presented in~\cite{Zhuang2017a,Zhuang2017b} but remains technically challenging.

Of course the simple advantage of using multiple entangled modes immediately helps with high loss
scenarios as now each mode can contain one heralded photon.
However, even with hundreds of modes, source brightness is limited to the sub-microwatt region.
Classical pulsed sources with thousands (or millions) of photons per pulse thus easily outperform
quantum illumination in most scenarios and being single-moded, narrowband filtering can be used to
reduce background.
Hence the use of (multi-moded) entangled sources (quantum illumination) needs to be motivated by
means other than enhanced signal-to-noise ratios (SNRs) or increased contrasts.

For LIDAR this justification can be found in the application of covert ranging or quantum
rangefinding.
In our protocol, entanglement is not used to improve the SNR compared to single-moded illumination
schemes.
However, one half of a bipartite entangled or two-mode squeezed state is always in a maximally mixed
state, indistinguishable from the state of a single mode of thermal background radiation.
The thermal photon statistics produced by one half of a two-mode squeezed state are a well known
result in quantum optics~\cite{Barnett1985,Yurke1987} which has also been confirmed
experimentally~\cite{Tapster1998,Blauensteiner2009}.
For completeness we provide a derivation in the supplementary material.

If a single spatial mode of spectrally multi-moded background can be replaced with one half of a
state produced in spontaneous parametric down-conversion (SPDC), efficient camouflaging can be
achieved, if the occupied spectral modes are identical to the ones replaced.
Of course, if the spectral density is not matched perfectly with the background light from the
region surrounding the rangefinder, a target can detect the probing beam by comparing the spectral
properties with neighbouring spatial modes.
Nevertheless, a broadband SPDC source will still perform much better than typically used pulsed
lasers with lower spectral bandwidths.
We quantify the minimum error probability of distinguishing the probing beam of our rangefinder from
neighbouring modes for different spectral overlaps in the supplementary material.

Such a broadband state can be tailored using quasi-phase
matching~\cite{Hum2007,Dosseva2014,Tambasco2016} in non-linear crystals such as periodically poled
potassium titanyl phosphate (ppKTP).
Careful engineering of the poling structure of these crystals can be used to emulate identical
spectral and photon statistical behaviour as a single spatial mode from thermal background radiation
emitted from the surroundings.
This means that a single spectral mode ($K=1$, with mode number $K$) will show the exact photon
statistics of a single mode of a truly thermal source in terms of their second order correlation
function $g^{(2)}$.
Similarly, $K>1$ spectral modes show the same behaviour as thermal background light
\begin{equation}
  \label{eq:1}
  g^{(2)}(0) \, = \, 1 + \frac{1}{K}.
\end{equation}
A proof of this behaviour is given in~\cite{Christ2011}.
Unfortunately, this effect also prohibits us from directly demonstrating the thermal statistics of
our probing beam experimentally.
Since our protocol probes the target with spectral mode number $K\gg 1$ narrowband spectral
filtering would be needed, which in turn would greatly diminish any signal from our broadband state.

\begin{figure}
  \centering
  \begin{tikzpicture}[scale=.7, transform shape]
    \input{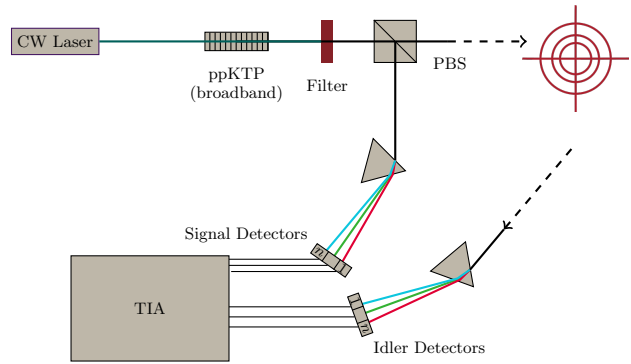}
  \end{tikzpicture}
  \caption{Schematic of a quantum rangefinder. The photon pairs produced in a continuous wave (CW)
    pumped ppKTP crystal are used to estimate the range to a target. While one photon of the pair is
    kept locally, the other is sent towards the target. The time difference of the two photons can
    be  used to estimate the distance since the idler photon is delayed by the time of flight to the
    target and back. The inverse energy correlation between signal and idler photons can be used to
    reduce background rates. The frequency bin of the idler photon can be predicted by measuring the
    frequency of the signal photon in $n$ channels. }\label{fig:schematic}
\end{figure}

Using a spectrally broadband state for illumination would typically result in high background
pollution of the signal, because a wider bandwidth of background has to be accepted by the
detectors.
Here the energy anti-correlation between signal and idler photons can be harnessed to achieve
background suppression (see figure~\ref{fig:schematic}).
If the locally kept photon is probed with respect to its colour, conservation of energy can be used
to predict the colour of its partner, if the frequency of the pump beam is known
\begin{equation}
  \label{eq:4}
  \hbar \omega_p \, = \, \hbar \omega_s + \hbar \omega_i,
\end{equation}
where $\hbar \omega_{\{p,s,i\}}$ denotes the energy of pump, signal and idler photons, respectively.
Using this fundamental principle in SPDC the background rate can be effectively filtered by
categorising photons and their partner into $n$ colour/frequency channels and only accepting events
that are constrained by energy conservation.


\section{Non-linear Crystal Design}\label{sec:non-linear-crystal}

Quasi phase-matching allows tailoring of the spectral properties of down-converted photons through
implementation of special poling periods $\Lambda$, engineered for specific applications.
This makes quasi phase-matching and the non-linear processes using it an extremely powerful tool
for quantum optics.
Engineering a poling period to emulate the behaviour of thermal background light, as is necessary
for efficient camouflaging, is a non-trivial task that involves sophisticated poling structures with
different periods.
For this purpose we developed a specialised software tool capable of predicting joint spectral
amplitudes (JSAs) of signal and idler photons from arbitrary poling structures.
Similar tools were developed by~\cite{Shalm,Dosseva2014}.

\begin{figure}
  \centering
  \begin{tikzpicture}
    \begin{axis}[xmin = 700, xmax = 949.8, ymin=701.527, ymax = 980.127, axis on top,
      xlabel={$\lambda_s$ [$\si{\nano\metre}$]},
      ylabel={$\lambda_i$ [$\si{\nano\metre}$]},
      ylabel style={yshift=-.2cm},
      xlabel style={yshift=.1cm},
      colorbar,
      colorbar style ={ylabel=arb. units, ylabel style={yshift=-2.5cm}},
      colormap={mycolormap}{color(0)=(ccb1) color(0.25)=(ccb2) color(0.5)=(ccb3) color(0.75)=(ccb4) color(1.0)=(ccb5)},
      width=.4\textwidth,
      xtick distance=50,
      ytick distance=50
      ]
      \addplot graphics[xmin = 700, xmax = 949.8, ymin=701.527, ymax = 980.127, includegraphics={bb=
      0 0 713 713}]
      {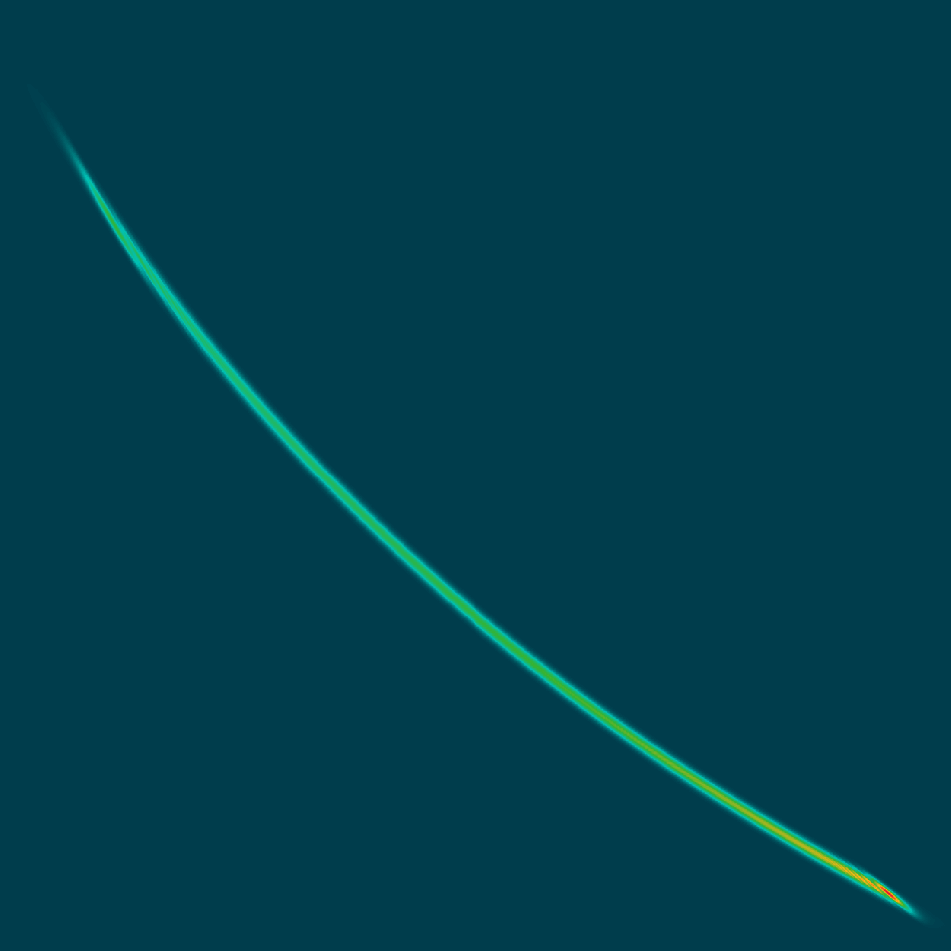};
    \end{axis}
  \end{tikzpicture}
  \caption{Resulting joint spectral intensity of a crystal designed with our specialised software.
    The custom poling of the crystal, comprising a linear chirp from $\Lambda =
    9\,\si{\micro\metre}$ to $\Lambda = 13 \,\si{\micro\metre}$, results in a phase-matching
    condition allowing for broadband photons between $700 \,\si{\nano\metre}$ and
    $950\,\si{\nano\metre}$ for both, signal and idler.}\label{fig:jsa01}
\end{figure}
Figure~\ref{fig:jsa01} shows the simulation of a poling structure designed with our software.
Introducing a chirp in the poling period $\Lambda$ from $9\,\si{\micro\metre}$ to
$13\,\si{\micro\metre}$ allows broadband type-II phase-matching which generates photons between
$700\,\si{\nano\metre}$ and $950\,\si{\nano\metre}$ wavelength from a $405 \, \si{\nano\metre}$ pump
laser.
These parameters where chosen since high power lasers at $405\,\si{\nano\metre}$ are ubiquitously
available while the longest down-converted wavelength of $950\,\si{\nano\metre}$ is still detectable
with off-the-shelf silicon avalanche photo diodes (APDs).
However, the non-linear material would allow for spectrally broader photons.

\section{Classical Signal-to-Noise Ratio Model}\label{sec:class-sign-noise}

While the ``quantum advantage'' in~\cite{Lloyd2008} was found by comparing the quantum Chernoff
bound~\cite{Audenaert2007} of a bi-partite state entangled over $n$ modes with the state of a
single photon.
We want to show here that we can infer a similar advantage by simple estimations of background and
coincidence rates as well as attenuation (optical gain $Q<1$) in the quantum channel of our
rangefinder.
This is still true when applying a linear detection scheme, where correlations between $n$ frequency
channels of signal and idler modes are considered.

Our model sets out to estimate the resulting number of coincident photon pairs $S$ in a
time-correlated histogram with bin width $\Delta t$ after an integration time $T$.
All the coincidences stemming from signal and idler modes occur in one time bin corresponding to the
distance of the target.
Hence, the time correlated histogram of photon events corresponds directly to a RADAR/LIDAR waveform
in classical systems.
We compare $S$ to the number of accidental coincidences $N$ appearing in such a histogram.
Events contributing to $N$  can be from a variety of different sources.
Besides background light our model also considers detector dark counts and imperfect heralding of
single photons.
Table~\ref{tab:SNRTerms} shows the different combinations of detector events that are considered in
our model and how they are denoted throughout this article.
\begin{table*}
  \centering
  \begin{tabular}{|l|c|c|c|c|}\hline
    \diagbox{Idler Mode}{Signal Mode}&
    Photon pair&
    Unpaired single photon&
    Background light &
    Dark count\\
    \hline
    Photon pair &
    $S$ &
    \catC\, $N_{s,c}$ &
    \catC\, $N_{B,c}$ &
    \catB\, $N_{d,c}$\\
    \hline
    Unpaired photon &
    \catC\, $N_{c,s}$  &
    \catC\, $N_{s,s}$  &
    \catC\, $N_{B,s}$ &
    \catB\, $N_{d,s}$\\
    \hline
    Background light &
    \catC\, $N_{c,B}$ &
    \catC\, $N_{s,B}$ &
    \catC\, $N_{B,B}$ &
    \catB\, $N_{d,B}$ \\
    \hline
    Dark count &
    \catB\, $N_{c,d}$ &
    \catB\, $N_{s,d}$ &
    \catB\, $N_{B,d}$ &
    \catA\, $N_{d,d}$\\
    \hline
  \end{tabular}
  \caption{Summary of all terms considered in our model
    and the symbols they are denoted with. Different combination of detector events can lead to
    different sources of noise $N$ and signal $S$. Red squares (\catA) mark
    terms proportional to the number of the frequency bins $n$, yellow triangles
    (\catB) label terms constant with the number of detectors and
    green circles (\protect\catC) mark terms inversely proportional to $n$.
  }\label{tab:SNRTerms}
\end{table*}
The signal-to-noise ratio (SNR) is then defined by the ratio of the number of photons in the LIDAR waveform's
coincidence peaks over the standard deviation of the noise and signal contributions combined
\begin{equation}
  \label{eq:2}
  \SNR \, = \, \frac{S}{\sqrt{S+N}},
\end{equation}
where $N$ comprises all noise terms from table~\ref{tab:SNRTerms} and Poissonian statistics are
assumed due to the discrete nature of photon counting.
For a full definition of all noise and signal terms please refer to the supplementary material.

Importantly, it is possible to categorise the noise terms into three different categories.
A first category $N_c$ (\catB) includes all noise terms that are constant with the number of
frequency channels $n$.
It comprises all combinations of background light and dark counts from table~\ref{tab:SNRTerms}.
A second category $N_p$ (\catA), which is proportional to the number of
frequency channels $n$, only considers dark counts in both arms.
Lastly, the category $N_i$ (\catC) is inversely proportional to the
number of channels.
Since this category contains all combinations of background and unpaired single photons it is
typically the largest contribution to the SNR and thereby guarantees the advantage from using the
frequency correlations inherent to the down-conversion process.
Our overall SNR model can thus be written as
\begin{equation}
  \label{eq:3}
  \SNR \, = \,  \frac{S}{\sqrt{S + N_c + n \cdot N_p + \frac{1}{n} \cdot N_i }}.
\end{equation}
From equation (\ref{eq:3}) the advantage gained by higher channel numbers is immediately visible and
is true while $n \leq n_\text{opt}$, with optimal channel number $n_\text{opt} = \sqrt{\frac{N_i}{N_p}}$.

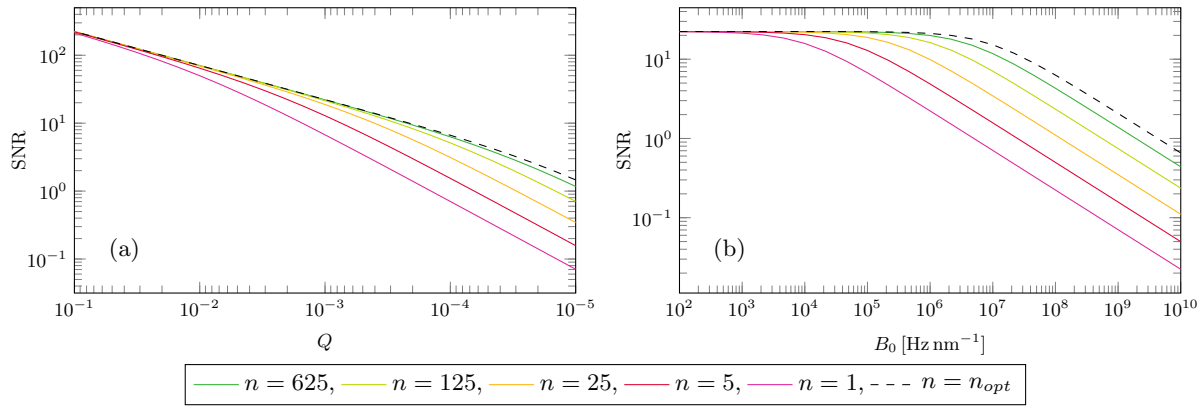
\begin{figure*}
  \centering
  \begin{tikzpicture}[baseline]
  \edef\CP{500000} \edef\LO{0.001} \edef\DT{0.0000000005} \edef\CD{500} \edef\B0{100000}
  \edef\DL{200}
  \begin{axis}[ xmode=log, ymode=log, xlabel={$Q$}, ylabel={SNR}, width=.46\textwidth,
    height=.3\textwidth, x dir = reverse, xmin = 0.00001, xmax = 0.1, legend columns = -1,
    legend to name = legendo, x label style={at={(axis description cs:0.5,0.02)},anchor=north,
      scale=0.8}, y label style={at={(axis description cs:0.1,0.5)},anchor=south, scale=0.8},
    tick label style={scale=0.8}, legend style={scale=0.8} ]
    \edef\SNR#1{(\CP*x)/sqrt(\CP*x+(\CD*\B0*\DL+\CD*x*\CP+\CP*\CD)*\DT+#1*(\CD*\CD*\DT)+(\CP*\B0*\DL*\DT)/#1)}
    \addplot[uobBrightGreen]
    expression[domain=\pgfkeysvalueof{/pgfplots/xmin}:\pgfkeysvalueof{/pgfplots/xmax}]
    {\SNR{625}}; \addlegendentry{$n=625$,} \addplot[uobBrightLime]
    expression[domain=\pgfkeysvalueof{/pgfplots/xmin}:\pgfkeysvalueof{/pgfplots/xmax}]
    {\SNR{125}}; \addlegendentry{$n=125$,} \addplot[uobBrightYellow]
    expression[domain=\pgfkeysvalueof{/pgfplots/xmin}:\pgfkeysvalueof{/pgfplots/xmax}]
    {\SNR{25}}; \addlegendentry{$n=25$,} \addplot[uobBrightRed]
    expression[domain=\pgfkeysvalueof{/pgfplots/xmin}:\pgfkeysvalueof{/pgfplots/xmax}]
    {\SNR{5}}; \addlegendentry{$n=5$,} \addplot[uobBrightPink]
    expression[domain=\pgfkeysvalueof{/pgfplots/xmin}:\pgfkeysvalueof{/pgfplots/xmax}]
    {\SNR{1}}; \addlegendentry{$n=1$,} \addplot[dashed]
    expression[domain=\pgfkeysvalueof{/pgfplots/xmin}:\pgfkeysvalueof{/pgfplots/xmax}]
    {\SNR{6324.55532034}}; \addlegendentry{$n=n_{opt}$}
    \node at (axis description cs:0.1,0.15) {(a)};
  \end{axis}
\end{tikzpicture}
\begin{tikzpicture}[baseline]
  \edef\CP{500000} \edef\LO{0.001} \edef\DT{0.0000000005} \edef\CD{500} \edef\B0{100000}
  \edef\DL{200}
  \begin{axis}[ xmode=log, ymode=log, xlabel={$B_0 \, [\si{\hertz\per\nano\metre}]$},
    ylabel={SNR}, width=.46\textwidth, height=.3\textwidth, xmin = 1e2, xmax = 1e10, x label
    style={at={(axis description cs:0.5,0.02)},anchor=north, scale=0.8}, y label
    style={at={(axis description cs:0.1,0.5)},anchor=south, scale=0.8}, tick label
    style={scale=0.8} ]
    \edef\SNR#1{(\CP*\LO)/sqrt(\CP*\LO+(\CD*x*\DL+\CD*\LO*\CP+\CP*\CD)*\DT+#1*(\CD*\CD*\DT)+(\CP*x*\DL*\DT)/#1)}
    \addplot[uobBrightGreen]
    expression[domain=\pgfkeysvalueof{/pgfplots/xmin}:\pgfkeysvalueof{/pgfplots/xmax}]
    {\SNR{625}}; \addplot[uobBrightLime]
    expression[domain=\pgfkeysvalueof{/pgfplots/xmin}:\pgfkeysvalueof{/pgfplots/xmax}]
    {\SNR{125}}; \addplot[uobBrightYellow]
    expression[domain=\pgfkeysvalueof{/pgfplots/xmin}:\pgfkeysvalueof{/pgfplots/xmax}]
    {\SNR{25}}; \addplot[uobBrightRed]
    expression[domain=\pgfkeysvalueof{/pgfplots/xmin}:\pgfkeysvalueof{/pgfplots/xmax}]
    {\SNR{5}}; \addplot[uobBrightPink]
    expression[domain=\pgfkeysvalueof{/pgfplots/xmin}:\pgfkeysvalueof{/pgfplots/xmax}]
    {\SNR{1}}; \addplot[dashed]
    expression[domain=\pgfkeysvalueof{/pgfplots/xmin}:\pgfkeysvalueof{/pgfplots/xmax}]
    {\SNR{6324.55532034}};
    \node at (axis description cs:0.1,0.15) {(b)};
  \end{axis}
\end{tikzpicture}

\ref{legendo}

  \caption{SNRs for different attenuations and background light levels. Plot (a) shows the
    estimated SNRs for different attenuations/optical gains $Q$ with a fixed background level of
    $B_0=100\,\si{\kilo\hertz}$.
    Different levels of background rates $B_0$ with a fixed optical gain $Q=10^{-3}$ are compared in plot (b).
    Both instances show the improvement of SNR with increasing channel numbers.}\label{fig:model}
\end{figure*}
This behaviour can also be observed in figure~\ref{fig:model}, where our model is plotted for
(a) different attenuations and (b) different background rates.
A clear advantage is gained by adding more channels while an effect of ``diminishing
returns'' becomes apparent with increasing channel numbers:
The increase of SNR between $n=1$ and $n=5$ is similar to the one between $n=25$ and $n=125$
reaching the optimal value at $n=n_{\text{opt}}$.
Typical numbers for $n_{\text{opt}}$ are $>10\,\si{\kilo}$ and occurs when the dark count rate of
the detectors surpasses the background rate per channel.
As with temporal correlations~\cite{Liu2019}, this advantage is purely due to classical frequency
correlations and could instead be implemented using a fast tunable laser.
However, the frequency correlations enable an additional advantage to the time correlations.
Early experimental realisations of quantum illumination used spatial correlations present in the
characteristic cone-shaped emission pattern from angle-phase matched SPDC
sources~\cite{Lopaeva2013}, which of course can also be emulated by classical physics.

\section{Experimental Verification}\label{sec:exper-verif}

The predictions made by our model are using Poissonian statistics only.
Thus, quantum mechanics is not needed to explain the advantage of increased SNR; this advantage can
be achieved by classical correlations.
However, quantum mechanics guarantees the covertness of the rangefinder.

Any practical implementation of our protocol will suffer from technical imperfections causing
additional noise in the system.
To prove the technical feasibility of our protocol, we here demonstrate it in a lab experiment.

\subsection{Experiment Setup}\label{sec:experiment-setup}

To verify our model for quantum rangefinding, we devised an experiment that can, at the same time,
examine multiple channel numbers $n \, \in \, \{ 1,2 \}$ under different background and loss conditions.

\begin{figure*}
  \centering
  \input{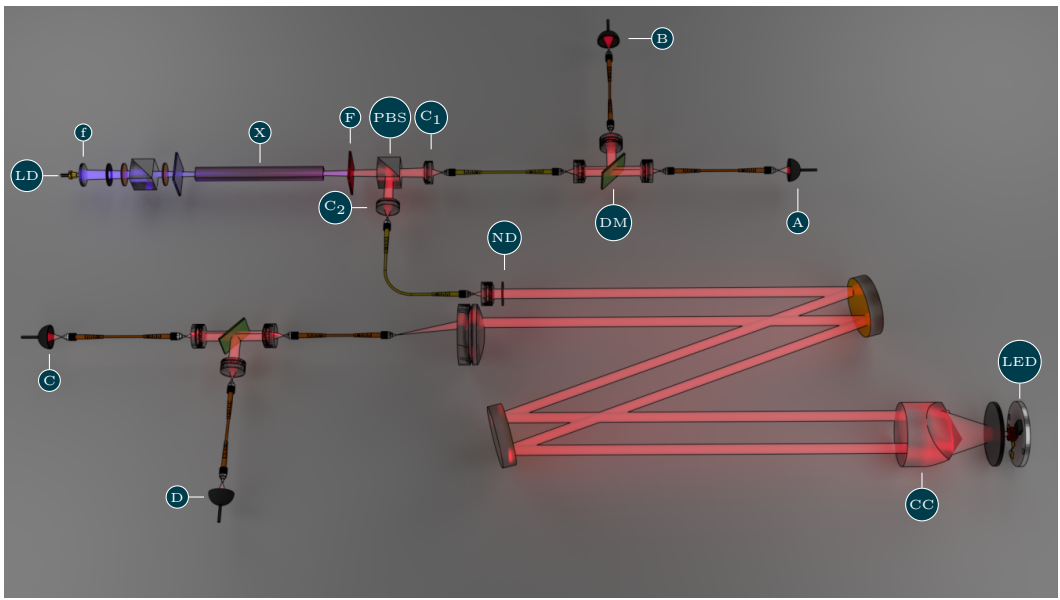}
  \caption{Complete setup as used to verify the SNR model. Light from an off-the-shelf
    $405\,\si{\nano\metre}$ laser diode (LD) is focused with an aspheric lens ($\text{f}$)
    into a non-linear crystal (X) after its  polarization is conditioned for the correct pump
    orientation for phase-matching. After removing the pump light with a long-pass filter
    ($\text{F}$) the  signal and idler photons are subsequently split on a polarizing beam-splitter
    (PBS) and collected into two single-mode fibres using the doublet couplers $\text{C}_{\text{1}}$
    and $\text{C}_{\text{2}}$. A channel number $n=2$ is implemented by using a dichroic mirror DM
    after coupling the locally kept photons back into free space.
    Depending on their wavelength these photons are then either guided to detector A or B. The other
    half of the photon pair is injected to a laser range consisting of two mirrors and a corner
    cube mirror (CC). Finally, the returning photon is coupled into an identical frequency splitting
    setup using a dichroic mirror and detectors C and D. Attenuations can be simulated by inserting
    a neutral density filter (ND) into the laser range, while back illumination of the corner cube
    with an LED can be used to simulate different background rates.}\label{fig:full-setup}
\end{figure*}
For this purpose the channel number $n=2$ was implemented using a frequency splitting setup realised
with a dichroic mirror for both signal and idler photons (see figure~\ref{fig:full-setup}).

With this setup we can simulate one frequency channel by combining all coincidences between the
four detectors
\begin{equation}
  \label{eq:5}
  C_{\text{total},n=1}=C_{AC}+C_{AD}+C_{BC}+C_{BD}
\end{equation}
or two frequency channels by combining only channels that should contain coincidences constrained by energy conservation
\begin{equation}
\label{eq:6}
    C_{\text{total},n=2}=C_{AD}+C_{BC}.
\end{equation}

All coincidence rates are calculated from a LIDAR waveform generated from a time-correlated
histogram of arrival times at the respective single photon detectors.
The time tagger used in this experiment is developed in house at the University of Bristol with a
resolution of $\approx 50\,\si{\pico\second}$ and a maximal event rate of $\approx
1\,\si{\mega\hertz}$~\cite{Nock2011}.

Additionally, custom electronics were engineered containing the current and temperature control for
the pump laser diode and temperature control for the crystal oven stabilised at $40\,\si{\celsius}$.
See supplementary material for details.


\subsection{Methods \& Results}\label{sec:methods--results}

First we verify the spectral properties of the custom poling structure of our down-conversion
crystal by separately measuring the signal and idler spectra on a single photon spectrometer (see
supplementary material) and by using the two-channel frequency splitting setup.
\begin{figure}
  \centering
  \begin{tikzpicture}[scale=.6, transform shape]
  \pgfmathsetmacro\scaleF{2}

  \pgfmathsetmacro\heightA{\scaleF*0.406771745476}
  \pgfmathsetmacro\heightB{\scaleF*0.0434325744308}
  \pgfmathsetmacro\heightC{\scaleF*0.529830706363}
  \pgfmathsetmacro\heightD{\scaleF*0.0199649737303}

  \pgfmathsetmacro\cAngleX{20}
  \pgfmathsetmacro\cAngleY{30}

  \pgfmathsetmacro\FOV{1.}

  \newcommand\setbVector[5]{
    \pgfmathsetmacro{\length}{#1*cos(\cAngleX)*cos(\cAngleY)+#2*sin(\cAngleX)*cos(\cAngleY)+#3*sin(\cAngleY)}
    \pgfmathsetmacro{#4}{(-#1*sin(\cAngleX)+#2*cos(\cAngleX))*\scaleF}
    \pgfmathsetmacro{#5}{(-#1*sin(\cAngleY)*cos(\cAngleX)-#2*sin(\cAngleY)*sin(\cAngleX)+#3*cos(\cAngleY))*\scaleF}
  }

  \setbVector{5}{0}{0}{\PAX}{\PAY}
  \setbVector{5}{5}{0}{\PBX}{\PBY}
  \setbVector{0}{5}{0}{\PCX}{\PCY}

  \pgfdeclareimage[height=2cm]{background}
  {./graphics/continuousJSA}

  \pgfmathsetmacro{\A}{(\PBX-\PAX)/2}
  \pgfmathsetmacro{\B}{(\PBY-\PAY)/2}
  \pgfmathsetmacro{\C}{(\PCX-\PBX)/2}
  \pgfmathsetmacro{\D}{(\PCY-\PBY)/2}
  \pgfmathsetmacro{\S}{-\PBX/2*1cm+0.08cm}
  \pgfmathsetmacro{\T}{\PBY/2*1cm-.4cm}

  \pgftext[at=\pgfpointorigin]{
    \pgflowlevel{\pgftransformcm{\A}{\B}{\C}{\D}{\pgfpoint{\S}{\T}}}
    \pgfuseimage{background}
  }

  \draw (0,0) -- (\PAX, \PAY) node[midway,above=.7cm,sloped] {Idler Wavelength [nm]} -- (\PBX,\PBY)
  node[midway,below=.5cm,sloped] {Signal Wavelength [nm]} -- (\PCX, \PCY) -- cycle;

  \foreach \w in {10,20,...,250}{
    \draw[white] ($(\PAX,\PAY)!\w/250!(\PBX,\PBY)$) -- ++ ($-0.02*(\PAX,\PAY)$);
  }
  \foreach \w in {9,19,...,279}{
    \draw[white] ($(\PAX,\PAY)!\w/279!(0,0)$) -- ++ ($0.015*(\PCX,\PCY)$);
  }

  \node[below] at ($(\PAX,\PAY)!10/250!(\PBX,\PBY)$) {710};
  \node[below] at ($(\PAX,\PAY)!240/250!(\PBX,\PBY)$) {940};
  \node[below] at ($(\PAX,\PAY)!140/250!(\PBX,\PBY)$) {840};

  \node[above, left] at ($(\PAX,\PAY)!9/279!(0,0)$) {710};
  \node[above, left] at ($(\PAX,\PAY)!269/279!(0,0)$) {970};
  \node[above, left] at ($(\PAX,\PAY)!149/279!(0,0)$) {850};

  \draw[white, thick, dashed] ($(\PAX,\PAY)!105/250!(\PBX,\PBY)$) node[below,black]{805}--
  ($(0,0)!105/250!(\PCX,\PCY)$);
  \draw[white, thick, dashed] ($(\PAX,\PAY)!104/279!(0,0)$) node[above, left, black]{805} --
  ($(\PBX,\PBY)!104/279!(\PCX,\PCY)$);

  \setbVector{1.1}{1.6}{0}{\SABAX}{\SABAY}
  \setbVector{2.1}{1.6}{0}{\SABBX}{\SABBY}
  \setbVector{2.1}{.6}{0}{\SABCX}{\SABCY}
  \setbVector{1.1}{1.6}{\heightA}{\SATAX}{\SATAY}
  \setbVector{2.1}{1.6}{\heightA}{\SATBX}{\SATBY}
  \setbVector{2.1}{.6}{\heightA}{\SATCX}{\SATCY}
  \setbVector{1.1}{.6}{\heightA}{\SATDX}{\SATDY}

  \draw[fill=gray, opacity=.65] (\SABAX, \SABAY) -- (\SABBX, \SABBY) -- (\SATBX, \SATBY) -- (\SATAX, \SATAY) -- cycle;
  \draw[fill=white, opacity=.65] (\SABCX, \SABCY) -- (\SABBX, \SABBY) -- (\SATBX, \SATBY) -- (\SATCX, \SATCY) -- cycle;
  \draw[fill=white, opacity=.65] (\SATAX, \SATAY) -- (\SATBX, \SATBY) --
  (\SATCX, \SATCY) node[midway, above, sloped, black] {$13.9\,\si{\kilo\hertz}$}  -- (\SATDX, \SATDY) -- cycle;

  \setbVector{3.5}{1.6}{0}{\SBBAX}{\SBBAY}
  \setbVector{4.5}{1.6}{0}{\SBBBX}{\SBBBY}
  \setbVector{4.5}{.6}{0}{\SBBCX}{\SBBCY}
  \setbVector{3.5}{1.6}{\heightB}{\SBTAX}{\SBTAY}
  \setbVector{4.5}{1.6}{\heightB}{\SBTBX}{\SBTBY}
  \setbVector{4.5}{.6}{\heightB}{\SBTCX}{\SBTCY}
  \setbVector{3.5}{.6}{\heightB}{\SBTDX}{\SBTDY}

  \draw[fill=gray, opacity=.65] (\SBBAX, \SBBAY) -- (\SBBBX, \SBBBY) -- (\SBTBX, \SBTBY) -- (\SBTAX, \SBTAY) -- cycle;
  \draw[fill=white, opacity=.65] (\SBBCX, \SBBCY) -- (\SBBBX, \SBBBY) -- (\SBTBX, \SBTBY) -- (\SBTCX, \SBTCY) -- cycle;
  \draw[fill=white, opacity=.65] (\SBTAX, \SBTAY) -- (\SBTBX, \SBTBY) --
  (\SBTCX, \SBTCY) node[midway, above, sloped, black] {$1.5\,\si{\kilo\hertz}$} -- (\SBTDX, \SBTDY) -- cycle;

  \setbVector{1.1}{4}{0}{\SDBAX}{\SDBAY}
  \setbVector{2.1}{4}{0}{\SDBBX}{\SDBBY}
  \setbVector{2.1}{3}{0}{\SDBCX}{\SDBCY}
  \setbVector{1.1}{4}{\heightD}{\SDTAX}{\SDTAY}
  \setbVector{2.1}{4}{\heightD}{\SDTBX}{\SDTBY}
  \setbVector{2.1}{3}{\heightD}{\SDTCX}{\SDTCY}
  \setbVector{1.1}{3}{\heightD}{\SDTDX}{\SDTDY}

  \draw[fill=gray, opacity=.65] (\SDBAX, \SDBAY) -- (\SDBBX, \SDBBY) -- (\SDTBX, \SDTBY) -- (\SDTAX, \SDTAY) -- cycle;
  \draw[fill=white, opacity=.65] (\SDBCX, \SDBCY) -- (\SDBBX, \SDBBY) -- (\SDTBX, \SDTBY) -- (\SDTCX, \SDTCY) -- cycle;
  \draw[fill=white, opacity=.65] (\SDTAX, \SDTAY) -- (\SDTBX, \SDTBY) --
  (\SDTCX, \SDTCY) -- (\SDTDX, \SDTDY) -- (\SDTAX, \SDTAY) node[midway, below, sloped, black] {$0.7\, \si{\kilo\hertz}$};

  \setbVector{3.5}{4}{0}{\SCBAX}{\SCBAY}
  \setbVector{4.5}{4}{0}{\SCBBX}{\SCBBY}
  \setbVector{4.5}{3}{0}{\SCBCX}{\SCBCY}
  \setbVector{3.5}{4}{\heightC}{\SCTAX}{\SCTAY}
  \setbVector{4.5}{4}{\heightC}{\SCTBX}{\SCTBY}
  \setbVector{4.5}{3}{\heightC}{\SCTCX}{\SCTCY}
  \setbVector{3.5}{3}{\heightC}{\SCTDX}{\SCTDY}

  \draw[fill=gray, opacity=.65] (\SCBAX, \SCBAY) -- (\SCBBX, \SCBBY) -- (\SCTBX, \SCTBY) -- (\SCTAX, \SCTAY) -- cycle;
  \draw[fill=white, opacity=.65] (\SCBCX, \SCBCY) -- (\SCBBX, \SCBBY) -- (\SCTBX, \SCTBY) -- (\SCTCX, \SCTCY) -- cycle;
  \draw[fill=white, opacity=.65] (\SCTAX, \SCTAY) -- (\SCTBX, \SCTBY) --
  (\SCTCX, \SCTCY) node[midway, above, sloped, black] {$18.2\,\si{\kilo\hertz}$} -- (\SCTDX, \SCTDY) -- cycle;

\end{tikzpicture}

  \caption{Coincidence rates between all four combinations of detectors using the two channel
    frequency splitting setup. The white lines show the cut-off/cut-on wavelength of the dichroic
    beam splitters. Almost all coincidences recorded are located in the off-diagonal contributions
    where energy conservation holds. Any accidental coincidences stemming from background light in
    the diagonal contributions are disregarded.}\label{fig:full-tomo}
\end{figure}
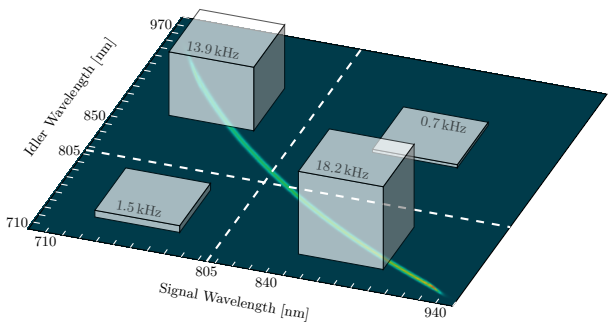
By measuring the coincidence rates between all four detectors we can confirm that all correlations
are contained within two of the four possible combinations of detectors
(figure~\ref{fig:full-tomo}).
This then justifies disregarding coincidences from the other two detector pairings, which arise
primarily from background light.

To compare cases of two and one-frequency channel the software of our time tagger is capable of
calculating time-correlated histograms between all four detectors in real-time.
These histograms are then summed up bin by bin, where all four histograms are considered to emulate
a one frequency channel solution~\eqref{eq:5} and the off-diagonal pairings represent the
two-channel setup~\eqref{eq:6}.
The resulting time-correlated histograms correspond to a LIDAR waveform as can be found in
conventional systems.
The peak position corresponds to the time-of-flight of the second photon to the target and
back and thus can be used to estimate the target distance.
Our time bin resolution in the histogram is chosen to be $750\,\si{\pico\second}$ and corresponds to
a range resolution of $\approx 10 \, \si{\centi\meter}$.
This is solely limited by the jitter of the chosen detectors of $600\,\si{\pico\second}$.
Contrary to passive, and thereby also covert, binocular/coincidence rangefinders, this resolution
is independent of the distance to the target, allowing high-precision measurments over far greater
range than the $3\,\si{\meter}$ demonstrated here.
Thus our protocol is advantageous over passive rangefinding techniques, especially since it is
easily adapted to better performing low jitter super-conducting photon detectors with jitter
$< 50\,\si{\pico\second}$.

\begin{figure}
  \centering
  \begin{tikzpicture}[baseline]
  \begin{axis}[
    width = .38\textwidth, height=.35\textwidth,
    xmin=0, xmax=60,
    ymin=0, ymax=350,
    xlabel={Delay [\si{\nano\second}]},
    ylabel={Coincidences per $1 \, \si{\second}$ and $750\, \si{\pico\second}$},
    x label style={at={(axis description cs:0.5,0.05)},anchor=north, scale=0.8},
    y label style={at={(axis description cs:0.075,0.5)},anchor=south, scale=0.8},
    tick label style={scale=0.8},
    xtick pos=left,
    ]
    \addplot[uobBrightRed, only marks, mark=x, domain=0:60, mark size=1pt, ultra thin] table
    {./data/histogram-109mA-9.90dB-allPairs_mod8.table};
    \addlegendentry{$n=1$}
    \addplot[uobBrightGreen, only marks, mark=x, domain=0:60, mark size=1pt, ultra thin] table
    {./data/histogram-109mA-9.90dB-goodPairs_mod8.table};
    \addlegendentry{$n=2$}
    \coordinate (meanStartTopGreen) at (axis cs:0,212);
    \coordinate (meanStartBottomGreen) at (axis cs:0,55.6);
    \coordinate (meanStartTopRed) at (axis cs:0,267.2);
    \coordinate (meanStartBottomRed) at (axis cs:0,111.9);
  \end{axis}
  \begin{axis}[
    width = .38\textwidth, height=.35\textwidth,
    xmin=-0.725, xmax=8.2687,
    ymin=0, ymax=350,
    ymajorticks = false,
    xlabel={Target Distance [\si{\meter}]},
    x label style={at={(axis description cs:0.5,1.17)},anchor=south, scale=0.8},
    tick label style={scale=0.8},
    xtick pos=right,
    axis x line*=top,
    axis y line=none
    ]
  \end{axis}
  \begin{axis}[
    xshift = .3\textwidth,
    width=.2\textwidth, height=.35\textwidth,
    ymin=0, ymax=350,
    xmin=-1, xmax=19,
    xlabel={Frequency},
    yticklabels={,,},
    xticklabels={,,},
    legend style = {anchor = north, at={(-1.7cm,-1.5cm), scale=0.8}},
    x label style={at={(axis description cs:0.5,0.05)},anchor=north, scale=0.8}
    ]
    \addplot[uobBrightGreen, fill, xbar, bar width=2.4, opacity=.8] table [x index = 1, y index = 0]
    {./data/histogram-109mA-9.90dB-GoodPeakDistr.table};
    \addplot[uobBrightRed, fill, xbar, bar width=2.4, opacity=.8] table [x index = 1, y index = 0]
    {./data/histogram-109mA-9.90dB-AllPeakDistr.table};
    \addplot[uobBrightGreen, fill, xbar, bar width=2.0, opacity=.8] table [x index = 1, y index = 0]
    {./data/histogram-109mA-9.90dB-GoodNoiseDistr.table};
    \addplot[uobBrightRed, fill, xbar, bar width=2.0, opacity=.8] table [x index = 1, y index = 0]
    {./data/histogram-109mA-9.90dB-AllNoiseDistr.table};
    \coordinate (meanStopTopGreen) at (axis cs:19,212);
    \coordinate (meanStopBottomGreen) at (axis cs:19,55.6);
    \coordinate (meanStopTopRed) at (axis cs:19,267.2);
    \coordinate (meanStopBottomRed) at (axis cs:19,111.9);
    \draw[uobBrightGreen, <->, thick] (axis cs:17,197.2) -- (axis cs:17,226.8);
    \draw[uobBrightGreen, <->, thick] (axis cs:17,47.3) -- (axis cs:17,63.9);
    \draw[uobBrightRed, <->, thick] (axis cs:17,283.8) -- (axis cs:17,250.6);
    \draw[uobBrightRed, <->, thick] (axis cs:17,100.5) -- (axis cs:17,123.3);
  \end{axis}
  \draw[uobBrightGreen, dashed] (meanStartTopGreen) -- (meanStopTopGreen);
  \draw[uobBrightGreen, dashed] (meanStartBottomGreen) -- (meanStopBottomGreen);
  \draw[uobBrightRed, dashed] (meanStartTopRed) -- (meanStopTopRed);
  \draw[uobBrightRed, dashed] (meanStartBottomRed) -- (meanStopBottomRed);
\end{tikzpicture}

  \caption{A typical measurement taken with the setup presented in figure~\ref{fig:full-setup}.
    The left plot shows an overlay of $600$ independent time-correlated histograms each recorded
    over $1\,\si{\second}$ integration time and with a bin width of $750\,\si{\pico\second}$ for
    both considered frequency channel numbers $n$. The resulting statistics for the signal peak and
    noise floor are extracted and plotted on the right side. Our time bin resolution is chosen to be
    $750\,\si{\pico\second}$, limited by the $600\,\si{\pico\second}$ jitter of the photon detectors
    used.}\label{fig:lidar-waveform}
\end{figure}
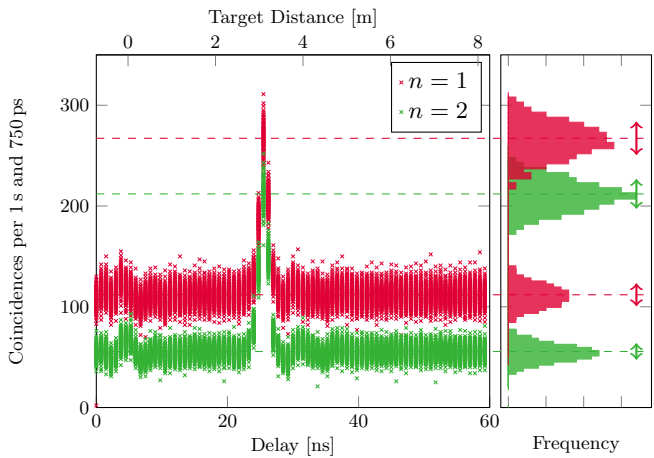

By recording $600$ of these waveforms we can estimate a distribution of signal peak heights above
the noise floor (figure~\ref{fig:lidar-waveform}).
These statistics are then used to calculate the SNR in our system.
Since our estimate of the SNR calculated here depends on both the mean and the variance of the
measured distributions, we are not able to directly infer measurement errors from the collected
data.
In order to quantify the accuracy of our measurements, we employ a fitting algorithm
(Levenberg-Marquardt~\cite{Levenberg1944,Marquardt1963}) to approximate the measured distributions
with a normal distribution.
We have chosen a normal distribution in this case, instead of a Poissonian distribution, to allow
for standard deviations that are independent of the mean value.
With this we can guarantee a fair comparison between model and experiment, in which technical noise
might lead to reduced SNRs.
The residual fitting error of the mean and standard deviation is then used to give the uncertainty
of our measurement.

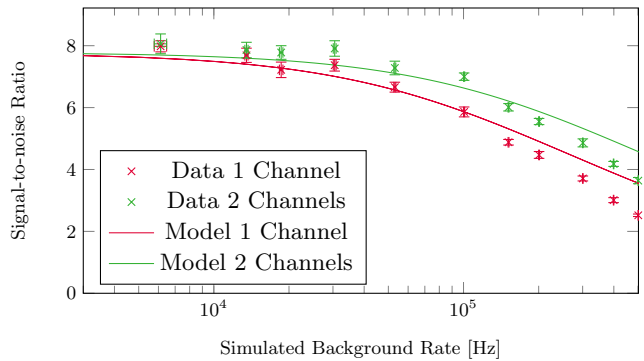
\begin{figure}
  \centering
  \begin{tikzpicture}
  \fill[white] (0,0) rectangle (7.4cm,4cm);
  \begin{axis}[
    xlabel = {Simulated Background Rate [\si{\hertz}]},
    ylabel = {Signal-to-noise Ratio},
    width=0.5\textwidth, height=0.3\textwidth,
    xmin=3000, xmax=502000,
    xmode=log,
    ymin=0,
    legend pos = south west,
    x label style={at={(axis description cs:0.5,0.0)},anchor=north, scale=0.8},
    y label style={at={(axis description cs:0.08,0.5)},anchor=south, scale=0.8},
    tick label style={scale=0.8}
    ]
    \addplot[uobBrightRed, only marks, mark=x,
    error bars/.cd,
    y dir=both,y explicit,
    x dir=both,x explicit]
    table[x error index=2, y error index=3]
    {data/SNR-vs-Background-18.1dB-Data-1Channel.table};
    \addlegendentry{Data 1 Channel}
    \addplot[uobBrightGreen, only marks, mark=x,
    error bars/.cd,
    y dir=both,y explicit,
    x dir=both,x explicit]
    table[x error index=2, y error index=3]
    {data/SNR-vs-Background-18.1dB-Data-2Channel.table};
    \addlegendentry{Data 2 Channels}
    \addplot[uobBrightRed, no markers]
    table{data/SNR-vs-Background-18.1dB-Model-1Channel.table};
    \addlegendentry{Model 1 Channel}
    \addplot[uobBrightGreen, no markers]
    table{data/SNR-vs-Background-18.1dB-Model-2Channel.table};
    \addlegendentry{Model 2 Channels}
  \end{axis}
\end{tikzpicture}
  \caption{Comparision between model and experiment for different background rates and an
    attenuation of $-18.1\,\si{\deci\bel}$. The signal-to-noise ratio is increased for the
    two-channel case and is accurately predicted by our model.}\label{fig:comparison}
\end{figure}
Figure~\ref{fig:comparison} depicts a comparison between our model and the data measured in the
experiment.
The absolute increase of SNR is predicted with good agreement while the absolute value deviates from
the model, especially for higher background rates.
This deviation is still well within a $<10\%$ error margin, or even within the error bars of
our experiment.
For high background rates the deviation is explained by saturating detectors/time tagging electronics.

Most importantly the advantage of using frequency anti-correlations, which occur naturally within
the SPDC process, shows a clear advantage by enabling the removal of unwanted background radiation.
This is possible despite the target being illuminated with a perfectly mixed/noisy state which is
not carrying any information in absence of its partner mode and is consequently undetectable.

\section{Conclusion \& Outlook}\label{sec:conclusion--outlook}

We have shown that multi-mode rangefinding can be used to develop a covert
rangefinding system operating at light levels significantly below daylight background and spectrally
and statistically indistinguishable from that background.

Our source exploits the energy anti-correlations between the broadband states produced in signal
and idler modes and we have constructed an SNR model equivalent to classical narrowband
illumination.
Although we only demonstrate a two-channel system here, our model predicts that employing a high
number of channels, such as $10^4$ would allow us to reach emission rates $>10^{12} \si{\hertz}$
(assuming $0.1$ pairs per $\Delta t \, \approx \, 1 \si{\nano\second}$ bin width) leading to return
loss tolerances of order $100\,\si{\deci\bel}$ enabling rangefinding of passive targets in high
background.
Such a source would still retain covertness even at higher pair generation rates due to the thermal
nature of the transmitted quantum state.

In this paper we present our work on covert or quantum rangefinding.
Taking advantage of frequency anti-correlations allows background noise to be removed from the
measurement, in addition to the background suppression achieved through temporal correlations of the
two-mode squeezed state, as was done in~\cite{Liu2019}.
While any SNR improving effects are purely due to classical correlations as in any earlier
demonstrated experiments, the covertness of one half of the two-mode squeezed state against thermal
background light is a useful and truly quantum resource that has thus far been overlooked by
previous work.
The possibility to use maximally mixed light, in a quantum state that carries no information (or
modulation), to illuminate a target and then to reconstruct position information using a measurement
realised with a linear detection scheme is only possible using the spectro-temporal entanglement of
the photon pair state.

\section*{Funding}
\label{sec:aknowledgements}

This work was supported by the EPSRC UK Quantum Technology Hub in Quantum Enhanced Imaging
(EP/M01326X/1).
SF was supported by the DSTL project ``Quantum illumination and the invisible rangefinder'' (DSTLX-1000091990)
and the Quantum Engineering Centre for Doctoral Training (EP/L015730/1).
JR would like to acknowledge EPSRC grant reference EP/M024458/1.

\section*{Open access data statement}
Open access data relating to this publication is available at the University of Bristol data
repository: \href{https://data.bris.ac.uk/data/}{https://data.bris.ac.uk/data/}

\bibliography{newLib}

\begin{thebibliography}{36}%
\makeatletter
\providecommand \@ifxundefined [1]{%
 \@ifx{#1\undefined}
}%
\providecommand \@ifnum [1]{%
 \ifnum #1\expandafter \@firstoftwo
 \else \expandafter \@secondoftwo
 \fi
}%
\providecommand \@ifx [1]{%
 \ifx #1\expandafter \@firstoftwo
 \else \expandafter \@secondoftwo
 \fi
}%
\providecommand \natexlab [1]{#1}%
\providecommand \enquote  [1]{``#1''}%
\providecommand \bibnamefont  [1]{#1}%
\providecommand \bibfnamefont [1]{#1}%
\providecommand \citenamefont [1]{#1}%
\providecommand \href@noop [0]{\@secondoftwo}%
\providecommand \href [0]{\begingroup \@sanitize@url \@href}%
\providecommand \@href[1]{\@@startlink{#1}\@@href}%
\providecommand \@@href[1]{\endgroup#1\@@endlink}%
\providecommand \@sanitize@url [0]{\catcode `\\12\catcode `\$12\catcode
  `\&12\catcode `\#12\catcode `\^12\catcode `\_12\catcode `\%12\relax}%
\providecommand \@@startlink[1]{}%
\providecommand \@@endlink[0]{}%
\providecommand \url  [0]{\begingroup\@sanitize@url \@url }%
\providecommand \@url [1]{\endgroup\@href {#1}{\urlprefix }}%
\providecommand \urlprefix  [0]{URL }%
\providecommand \Eprint [0]{\href }%
\providecommand \doibase [0]{http://dx.doi.org/}%
\providecommand \selectlanguage [0]{\@gobble}%
\providecommand \bibinfo  [0]{\@secondoftwo}%
\providecommand \bibfield  [0]{\@secondoftwo}%
\providecommand \translation [1]{[#1]}%
\providecommand \BibitemOpen [0]{}%
\providecommand \bibitemStop [0]{}%
\providecommand \bibitemNoStop [0]{.\EOS\space}%
\providecommand \EOS [0]{\spacefactor3000\relax}%
\providecommand \BibitemShut  [1]{\csname bibitem#1\endcsname}%
\let\auto@bib@innerbib\@empty
\bibitem [{\citenamefont {O'Brien}(2007)}]{O'Brien2007}%
  \BibitemOpen
  \bibfield  {author} {\bibinfo {author} {\bibfnamefont {J.~L.}\ \bibnamefont
  {O'Brien}},\ }\href {\doibase 10.1126/science.1142892} {\bibfield  {journal}
  {\bibinfo  {journal} {Science (New York, N.Y.)}\ }\textbf {\bibinfo {volume}
  {318}},\ \bibinfo {pages} {1567} (\bibinfo {year} {2007})}\BibitemShut
  {NoStop}%
\bibitem [{\citenamefont {O'Brien}\ \emph {et~al.}(2003)\citenamefont
  {O'Brien}, \citenamefont {Pryde}, \citenamefont {White}, \citenamefont
  {Ralph},\ and\ \citenamefont {Branning}}]{O'Brien2003}%
  \BibitemOpen
  \bibfield  {author} {\bibinfo {author} {\bibfnamefont {J.~L.}\ \bibnamefont
  {O'Brien}}, \bibinfo {author} {\bibfnamefont {G.~J.}\ \bibnamefont {Pryde}},
  \bibinfo {author} {\bibfnamefont {A.~G.}\ \bibnamefont {White}}, \bibinfo
  {author} {\bibfnamefont {T.~C.}\ \bibnamefont {Ralph}}, \ and\ \bibinfo
  {author} {\bibfnamefont {D.}~\bibnamefont {Branning}},\ }\href {\doibase
  10.1038/nature02054} {\bibfield  {journal} {\bibinfo  {journal} {Nature}\
  }\textbf {\bibinfo {volume} {426}},\ \bibinfo {pages} {264} (\bibinfo {year}
  {2003})}\BibitemShut {NoStop}%
\bibitem [{\citenamefont {O'Brien}\ \emph {et~al.}(2009)\citenamefont
  {O'Brien}, \citenamefont {Furusawa},\ and\ \citenamefont
  {Vu{\v{c}}kovi{\'{c}}}}]{O'Brien2009}%
  \BibitemOpen
  \bibfield  {author} {\bibinfo {author} {\bibfnamefont {J.~L.}\ \bibnamefont
  {O'Brien}}, \bibinfo {author} {\bibfnamefont {A.}~\bibnamefont {Furusawa}}, \
  and\ \bibinfo {author} {\bibfnamefont {J.}~\bibnamefont
  {Vu{\v{c}}kovi{\'{c}}}},\ }\href {\doibase 10.1038/nphoton.2009.229}
  {\bibfield  {journal} {\bibinfo  {journal} {Nature Photonics}\ }\textbf
  {\bibinfo {volume} {3}},\ \bibinfo {pages} {687} (\bibinfo {year}
  {2009})}\BibitemShut {NoStop}%
\bibitem [{\citenamefont {Pirandola}\ \emph {et~al.}(2018)\citenamefont
  {Pirandola}, \citenamefont {Bardhan}, \citenamefont {Gehring}, \citenamefont
  {Weedbrook},\ and\ \citenamefont {Lloyd}}]{Pirandola2018}%
  \BibitemOpen
  \bibfield  {author} {\bibinfo {author} {\bibfnamefont {S.}~\bibnamefont
  {Pirandola}}, \bibinfo {author} {\bibfnamefont {B.~R.}\ \bibnamefont
  {Bardhan}}, \bibinfo {author} {\bibfnamefont {T.}~\bibnamefont {Gehring}},
  \bibinfo {author} {\bibfnamefont {C.}~\bibnamefont {Weedbrook}}, \ and\
  \bibinfo {author} {\bibfnamefont {S.}~\bibnamefont {Lloyd}},\ }\href
  {\doibase 10.1038/s41566-018-0301-6} {\bibfield  {journal} {\bibinfo
  {journal} {Nature Photonics}\ }\textbf {\bibinfo {volume} {12}},\ \bibinfo
  {pages} {724} (\bibinfo {year} {2018})},\ \Eprint
  {http://arxiv.org/abs/1811.01969} {arXiv:1811.01969} \BibitemShut {NoStop}%
\bibitem [{\citenamefont {Genovese}(2016)}]{Genovese2016}%
  \BibitemOpen
  \bibfield  {author} {\bibinfo {author} {\bibfnamefont {M.}~\bibnamefont
  {Genovese}},\ }\href {\doibase 10.1088/2040-8978/18/7/073002} {\bibfield
  {journal} {\bibinfo  {journal} {Journal of Optics}\ }\textbf {\bibinfo
  {volume} {18}} (\bibinfo {year} {2016}),\ 10.1088/2040-8978/18/7/073002},\
  \Eprint {http://arxiv.org/abs/1601.06066} {arXiv:1601.06066} \BibitemShut
  {NoStop}%
\bibitem [{\citenamefont {Adesso}\ \emph {et~al.}(2016)\citenamefont {Adesso},
  \citenamefont {Bromley},\ and\ \citenamefont {Cianciaruso}}]{Adesso2016}%
  \BibitemOpen
  \bibfield  {author} {\bibinfo {author} {\bibfnamefont {G.}~\bibnamefont
  {Adesso}}, \bibinfo {author} {\bibfnamefont {T.~R.}\ \bibnamefont {Bromley}},
  \ and\ \bibinfo {author} {\bibfnamefont {M.}~\bibnamefont {Cianciaruso}},\
  }\href {\doibase 10.1088/1751-8113/49/47/473001} {\bibfield  {journal}
  {\bibinfo  {journal} {Journal of Physics A: Mathematical and Theoretical}\
  }\textbf {\bibinfo {volume} {49}} (\bibinfo {year} {2016}),\
  10.1088/1751-8113/49/47/473001},\ \Eprint {http://arxiv.org/abs/1605.00806}
  {arXiv:1605.00806} \BibitemShut {NoStop}%
\bibitem [{\citenamefont {Moreau}\ \emph {et~al.}(2017)\citenamefont {Moreau},
  \citenamefont {Sabines-Chesterking}, \citenamefont {Whittaker}, \citenamefont
  {Joshi}, \citenamefont {Birchall}, \citenamefont {McMillan}, \citenamefont
  {Rarity},\ and\ \citenamefont {Matthews}}]{Moreau2017}%
  \BibitemOpen
  \bibfield  {author} {\bibinfo {author} {\bibfnamefont {P.~A.}\ \bibnamefont
  {Moreau}}, \bibinfo {author} {\bibfnamefont {J.}~\bibnamefont
  {Sabines-Chesterking}}, \bibinfo {author} {\bibfnamefont {R.}~\bibnamefont
  {Whittaker}}, \bibinfo {author} {\bibfnamefont {S.~K.}\ \bibnamefont
  {Joshi}}, \bibinfo {author} {\bibfnamefont {P.~M.}\ \bibnamefont {Birchall}},
  \bibinfo {author} {\bibfnamefont {A.}~\bibnamefont {McMillan}}, \bibinfo
  {author} {\bibfnamefont {J.~G.}\ \bibnamefont {Rarity}}, \ and\ \bibinfo
  {author} {\bibfnamefont {J.~C.}\ \bibnamefont {Matthews}},\ }\href {\doibase
  10.1038/s41598-017-06545-w} {\bibfield  {journal} {\bibinfo  {journal}
  {Scientific Reports}\ }\textbf {\bibinfo {volume} {7}},\ \bibinfo {pages} {1}
  (\bibinfo {year} {2017})},\ \Eprint {http://arxiv.org/abs/1611.07871}
  {arXiv:1611.07871} \BibitemShut {NoStop}%
\bibitem [{\citenamefont {Whittaker}\ \emph {et~al.}(2017)\citenamefont
  {Whittaker}, \citenamefont {Erven}, \citenamefont {Neville}, \citenamefont
  {Berry}, \citenamefont {O'Brien}, \citenamefont {Cable},\ and\ \citenamefont
  {Matthews}}]{Whittaker2017}%
  \BibitemOpen
  \bibfield  {author} {\bibinfo {author} {\bibfnamefont {R.}~\bibnamefont
  {Whittaker}}, \bibinfo {author} {\bibfnamefont {C.}~\bibnamefont {Erven}},
  \bibinfo {author} {\bibfnamefont {A.}~\bibnamefont {Neville}}, \bibinfo
  {author} {\bibfnamefont {M.}~\bibnamefont {Berry}}, \bibinfo {author}
  {\bibfnamefont {J.~L.}\ \bibnamefont {O'Brien}}, \bibinfo {author}
  {\bibfnamefont {H.}~\bibnamefont {Cable}}, \ and\ \bibinfo {author}
  {\bibfnamefont {J.~C.~F.}\ \bibnamefont {Matthews}},\ }\href {\doibase
  10.1088/1367-2630/aa5512} {\bibfield  {journal} {\bibinfo  {journal} {New
  Journal of Physics}\ }\textbf {\bibinfo {volume} {19}},\ \bibinfo {pages}
  {023013} (\bibinfo {year} {2017})}\BibitemShut {NoStop}%
\bibitem [{\citenamefont {Lopaeva}\ \emph {et~al.}(2013)\citenamefont
  {Lopaeva}, \citenamefont {{Ruo Berchera}}, \citenamefont {Degiovanni},
  \citenamefont {Olivares}, \citenamefont {Brida},\ and\ \citenamefont
  {Genovese}}]{Lopaeva2013}%
  \BibitemOpen
  \bibfield  {author} {\bibinfo {author} {\bibfnamefont {E.~D.}\ \bibnamefont
  {Lopaeva}}, \bibinfo {author} {\bibfnamefont {I.}~\bibnamefont {{Ruo
  Berchera}}}, \bibinfo {author} {\bibfnamefont {I.~P.}\ \bibnamefont
  {Degiovanni}}, \bibinfo {author} {\bibfnamefont {S.}~\bibnamefont
  {Olivares}}, \bibinfo {author} {\bibfnamefont {G.}~\bibnamefont {Brida}}, \
  and\ \bibinfo {author} {\bibfnamefont {M.}~\bibnamefont {Genovese}},\ }\href
  {\doibase 10.1103/PhysRevLett.110.153603} {\bibfield  {journal} {\bibinfo
  {journal} {Physical Review Letters}\ }\textbf {\bibinfo {volume} {110}},\
  \bibinfo {pages} {1} (\bibinfo {year} {2013})},\ \Eprint
  {http://arxiv.org/abs/arXiv:1303.4304v1} {arXiv:arXiv:1303.4304v1}
  \BibitemShut {NoStop}%
\bibitem [{\citenamefont {Shadbolt}\ \emph {et~al.}(2012)\citenamefont
  {Shadbolt}, \citenamefont {Verde}, \citenamefont {Peruzzo}, \citenamefont
  {Politi}, \citenamefont {Laing}, \citenamefont {Lobino}, \citenamefont
  {Matthews}, \citenamefont {Thompson},\ and\ \citenamefont
  {O'Brien}}]{Shadbolt2011}%
  \BibitemOpen
  \bibfield  {author} {\bibinfo {author} {\bibfnamefont {P.~J.}\ \bibnamefont
  {Shadbolt}}, \bibinfo {author} {\bibfnamefont {M.~R.}\ \bibnamefont {Verde}},
  \bibinfo {author} {\bibfnamefont {A.}~\bibnamefont {Peruzzo}}, \bibinfo
  {author} {\bibfnamefont {A.}~\bibnamefont {Politi}}, \bibinfo {author}
  {\bibfnamefont {A.}~\bibnamefont {Laing}}, \bibinfo {author} {\bibfnamefont
  {M.}~\bibnamefont {Lobino}}, \bibinfo {author} {\bibfnamefont {J.~C.}\
  \bibnamefont {Matthews}}, \bibinfo {author} {\bibfnamefont {M.~G.}\
  \bibnamefont {Thompson}}, \ and\ \bibinfo {author} {\bibfnamefont {J.~L.}\
  \bibnamefont {O'Brien}},\ }\href {\doibase 10.1038/nphoton.2011.283}
  {\bibfield  {journal} {\bibinfo  {journal} {Nature Photonics}\ }\textbf
  {\bibinfo {volume} {6}},\ \bibinfo {pages} {45} (\bibinfo {year} {2012})},\
  \Eprint {http://arxiv.org/abs/1108.3309} {arXiv:1108.3309} \BibitemShut
  {NoStop}%
\bibitem [{\citenamefont {Seward}\ \emph {et~al.}(1991)\citenamefont {Seward},
  \citenamefont {Tapster}, \citenamefont {Walker},\ and\ \citenamefont
  {Rarity}}]{Seward1991}%
  \BibitemOpen
  \bibfield  {author} {\bibinfo {author} {\bibfnamefont {S.~F.}\ \bibnamefont
  {Seward}}, \bibinfo {author} {\bibfnamefont {P.~R.}\ \bibnamefont {Tapster}},
  \bibinfo {author} {\bibfnamefont {J.~G.}\ \bibnamefont {Walker}}, \ and\
  \bibinfo {author} {\bibfnamefont {J.~G.}\ \bibnamefont {Rarity}},\ }\href
  {\doibase 10.1088/0954-8998/3/4/002} {\bibfield  {journal} {\bibinfo
  {journal} {Quantum Optics: Journal of the European Optical Society Part B}\
  }\textbf {\bibinfo {volume} {3}},\ \bibinfo {pages} {201} (\bibinfo {year}
  {1991})}\BibitemShut {NoStop}%
\bibitem [{\citenamefont {Rarity}\ \emph {et~al.}(1990)\citenamefont {Rarity},
  \citenamefont {Tapster}, \citenamefont {Walker},\ and\ \citenamefont
  {Seward}}]{Rarity1990}%
  \BibitemOpen
  \bibfield  {author} {\bibinfo {author} {\bibfnamefont {J.~G.}\ \bibnamefont
  {Rarity}}, \bibinfo {author} {\bibfnamefont {P.~R.}\ \bibnamefont {Tapster}},
  \bibinfo {author} {\bibfnamefont {J.~G.}\ \bibnamefont {Walker}}, \ and\
  \bibinfo {author} {\bibfnamefont {S.}~\bibnamefont {Seward}},\ }\href@noop {}
  {\bibfield  {journal} {\bibinfo  {journal} {Applied optics}\ }\textbf
  {\bibinfo {volume} {29}},\ \bibinfo {pages} {2939} (\bibinfo {year}
  {1990})}\BibitemShut {NoStop}%
\bibitem [{\citenamefont {Liu}\ \emph {et~al.}(2019)\citenamefont {Liu},
  \citenamefont {Giovannini}, \citenamefont {He}, \citenamefont {England},
  \citenamefont {Sussman}, \citenamefont {Balaji},\ and\ \citenamefont
  {Helmy}}]{Liu2019}%
  \BibitemOpen
  \bibfield  {author} {\bibinfo {author} {\bibfnamefont {H.}~\bibnamefont
  {Liu}}, \bibinfo {author} {\bibfnamefont {D.}~\bibnamefont {Giovannini}},
  \bibinfo {author} {\bibfnamefont {H.}~\bibnamefont {He}}, \bibinfo {author}
  {\bibfnamefont {D.}~\bibnamefont {England}}, \bibinfo {author} {\bibfnamefont
  {B.~J.}\ \bibnamefont {Sussman}}, \bibinfo {author} {\bibfnamefont
  {B.}~\bibnamefont {Balaji}}, \ and\ \bibinfo {author} {\bibfnamefont {A.~S.}\
  \bibnamefont {Helmy}},\ }\href {\doibase 10.1364/optica.6.001349} {\bibfield
  {journal} {\bibinfo  {journal} {Optica}\ }\textbf {\bibinfo {volume} {6}},\
  \bibinfo {pages} {1349} (\bibinfo {year} {2019})},\ \Eprint
  {http://arxiv.org/abs/2004.06754} {arXiv:2004.06754} \BibitemShut {NoStop}%
\bibitem [{\citenamefont {Lloyd}(2008)}]{Lloyd2008}%
  \BibitemOpen
  \bibfield  {author} {\bibinfo {author} {\bibfnamefont {S.}~\bibnamefont
  {Lloyd}},\ }\href {\doibase 10.1126/science.1160627} {\bibfield  {journal}
  {\bibinfo  {journal} {Science}\ }\textbf {\bibinfo {volume} {321}},\ \bibinfo
  {pages} {1463} (\bibinfo {year} {2008})},\ \Eprint
  {http://arxiv.org/abs/0803.2022} {arXiv:0803.2022} \BibitemShut {NoStop}%
\bibitem [{\citenamefont {Shapiro}\ and\ \citenamefont
  {Lloyd}(2009)}]{Shapiro2009}%
  \BibitemOpen
  \bibfield  {author} {\bibinfo {author} {\bibfnamefont {J.~H.}\ \bibnamefont
  {Shapiro}}\ and\ \bibinfo {author} {\bibfnamefont {S.}~\bibnamefont
  {Lloyd}},\ }\href {\doibase 10.1088/1367-2630/11/6/063045} {\bibfield
  {journal} {\bibinfo  {journal} {New Journal of Physics}\ }\textbf {\bibinfo
  {volume} {11}},\ \bibinfo {pages} {063045} (\bibinfo {year} {2009})},\
  \Eprint {http://arxiv.org/abs/0902.0986} {arXiv:0902.0986} \BibitemShut
  {NoStop}%
\bibitem [{\citenamefont {Zhang}\ \emph {et~al.}(2014)\citenamefont {Zhang},
  \citenamefont {Guo}, \citenamefont {Bao}, \citenamefont {Shi}, \citenamefont
  {Jin}, \citenamefont {Zou},\ and\ \citenamefont {Guo}}]{Zhang2014}%
  \BibitemOpen
  \bibfield  {author} {\bibinfo {author} {\bibfnamefont {S.}~\bibnamefont
  {Zhang}}, \bibinfo {author} {\bibfnamefont {J.}~\bibnamefont {Guo}}, \bibinfo
  {author} {\bibfnamefont {W.}~\bibnamefont {Bao}}, \bibinfo {author}
  {\bibfnamefont {J.}~\bibnamefont {Shi}}, \bibinfo {author} {\bibfnamefont
  {C.}~\bibnamefont {Jin}}, \bibinfo {author} {\bibfnamefont {X.}~\bibnamefont
  {Zou}}, \ and\ \bibinfo {author} {\bibfnamefont {G.}~\bibnamefont {Guo}},\
  }\href {\doibase 10.1103/PhysRevA.89.062309} {\bibfield  {journal} {\bibinfo
  {journal} {Physical Review A}\ }\textbf {\bibinfo {volume} {89}},\ \bibinfo
  {pages} {062309} (\bibinfo {year} {2014})}\BibitemShut {NoStop}%
\bibitem [{\citenamefont {Tan}\ \emph {et~al.}(2008)\citenamefont {Tan},
  \citenamefont {Erkmen}, \citenamefont {Giovannetti}, \citenamefont {Guha},
  \citenamefont {Lloyd}, \citenamefont {Maccone}, \citenamefont {Pirandola},\
  and\ \citenamefont {Shapiro}}]{Tan2008}%
  \BibitemOpen
  \bibfield  {author} {\bibinfo {author} {\bibfnamefont {S.~H.}\ \bibnamefont
  {Tan}}, \bibinfo {author} {\bibfnamefont {B.~I.}\ \bibnamefont {Erkmen}},
  \bibinfo {author} {\bibfnamefont {V.}~\bibnamefont {Giovannetti}}, \bibinfo
  {author} {\bibfnamefont {S.}~\bibnamefont {Guha}}, \bibinfo {author}
  {\bibfnamefont {S.}~\bibnamefont {Lloyd}}, \bibinfo {author} {\bibfnamefont
  {L.}~\bibnamefont {Maccone}}, \bibinfo {author} {\bibfnamefont
  {S.}~\bibnamefont {Pirandola}}, \ and\ \bibinfo {author} {\bibfnamefont
  {J.~H.}\ \bibnamefont {Shapiro}},\ }\href {\doibase
  10.1103/PhysRevLett.101.253601} {\bibfield  {journal} {\bibinfo  {journal}
  {Physical Review Letters}\ }\textbf {\bibinfo {volume} {101}},\ \bibinfo
  {pages} {1} (\bibinfo {year} {2008})},\ \Eprint
  {http://arxiv.org/abs/0810.0534} {arXiv:0810.0534} \BibitemShut {NoStop}%
\bibitem [{\citenamefont {Shapiro}(2020)}]{Shapiro2020}%
  \BibitemOpen
  \bibfield  {author} {\bibinfo {author} {\bibfnamefont {J.~H.}\ \bibnamefont
  {Shapiro}},\ }\href {\doibase 10.1109/MAES.2019.2957870} {\bibfield
  {journal} {\bibinfo  {journal} {IEEE Aerospace and Electronic Systems
  Magazine}\ }\textbf {\bibinfo {volume} {35}},\ \bibinfo {pages} {8} (\bibinfo
  {year} {2020})}\BibitemShut {NoStop}%
\bibitem [{\citenamefont {Guha}\ and\ \citenamefont {Erkmen}(2009)}]{Guha2009}%
  \BibitemOpen
  \bibfield  {author} {\bibinfo {author} {\bibfnamefont {S.}~\bibnamefont
  {Guha}}\ and\ \bibinfo {author} {\bibfnamefont {B.~I.}\ \bibnamefont
  {Erkmen}},\ }\href {\doibase 10.1103/PhysRevA.80.052310} {\bibfield
  {journal} {\bibinfo  {journal} {Physical Review A - Atomic, Molecular, and
  Optical Physics}\ }\textbf {\bibinfo {volume} {80}},\ \bibinfo {pages} {1}
  (\bibinfo {year} {2009})},\ \Eprint {http://arxiv.org/abs/0911.0950}
  {arXiv:0911.0950} \BibitemShut {NoStop}%
\bibitem [{\citenamefont {Zhang}\ \emph {et~al.}(2015)\citenamefont {Zhang},
  \citenamefont {Mouradian}, \citenamefont {Wong},\ and\ \citenamefont
  {Shapiro}}]{Zheshen2015}%
  \BibitemOpen
  \bibfield  {author} {\bibinfo {author} {\bibfnamefont {Z.}~\bibnamefont
  {Zhang}}, \bibinfo {author} {\bibfnamefont {S.}~\bibnamefont {Mouradian}},
  \bibinfo {author} {\bibfnamefont {F.~N.~C.}\ \bibnamefont {Wong}}, \ and\
  \bibinfo {author} {\bibfnamefont {J.~H.}\ \bibnamefont {Shapiro}},\ }\href
  {\doibase 10.1103/PhysRevLett.114.110506} {\bibfield  {journal} {\bibinfo
  {journal} {Physical Review Letters}\ }\textbf {\bibinfo {volume} {114}},\
  \bibinfo {pages} {110506} (\bibinfo {year} {2015})},\ \Eprint
  {http://arxiv.org/abs/1411.5969} {arXiv:1411.5969} \BibitemShut {NoStop}%
\bibitem [{\citenamefont {Zhuang}\ \emph
  {et~al.}(2017{\natexlab{a}})\citenamefont {Zhuang}, \citenamefont {Zhang},\
  and\ \citenamefont {Shapiro}}]{Zhuang2017a}%
  \BibitemOpen
  \bibfield  {author} {\bibinfo {author} {\bibfnamefont {Q.}~\bibnamefont
  {Zhuang}}, \bibinfo {author} {\bibfnamefont {Z.}~\bibnamefont {Zhang}}, \
  and\ \bibinfo {author} {\bibfnamefont {J.~H.}\ \bibnamefont {Shapiro}},\
  }\href {\doibase 10.1103/PhysRevLett.118.040801} {\bibfield  {journal}
  {\bibinfo  {journal} {Physical Review Letters}\ }\textbf {\bibinfo {volume}
  {118}},\ \bibinfo {pages} {040801} (\bibinfo {year} {2017}{\natexlab{a}})},\
  \Eprint {http://arxiv.org/abs/1609.01968} {arXiv:1609.01968} \BibitemShut
  {NoStop}%
\bibitem [{\citenamefont {Zhuang}\ \emph
  {et~al.}(2017{\natexlab{b}})\citenamefont {Zhuang}, \citenamefont {Zhang},\
  and\ \citenamefont {Shapiro}}]{Zhuang2017b}%
  \BibitemOpen
  \bibfield  {author} {\bibinfo {author} {\bibfnamefont {Q.}~\bibnamefont
  {Zhuang}}, \bibinfo {author} {\bibfnamefont {Z.}~\bibnamefont {Zhang}}, \
  and\ \bibinfo {author} {\bibfnamefont {J.~H.}\ \bibnamefont {Shapiro}},\
  }\href {\doibase 10.1103/PhysRevA.96.020302} {\bibfield  {journal} {\bibinfo
  {journal} {Physical Review A}\ }\textbf {\bibinfo {volume} {96}},\ \bibinfo
  {pages} {1} (\bibinfo {year} {2017}{\natexlab{b}})},\ \Eprint
  {http://arxiv.org/abs/1706.05561} {arXiv:1706.05561} \BibitemShut {NoStop}%
\bibitem [{\citenamefont {Barnett}\ and\ \citenamefont
  {Knight}(1985)}]{Barnett1985}%
  \BibitemOpen
  \bibfield  {author} {\bibinfo {author} {\bibfnamefont {S.~M.}\ \bibnamefont
  {Barnett}}\ and\ \bibinfo {author} {\bibfnamefont {P.~L.}\ \bibnamefont
  {Knight}},\ }\href {\doibase 10.1364/josab.2.000467} {\bibfield  {journal}
  {\bibinfo  {journal} {Journal of the Optical Society of America B}\ }\textbf
  {\bibinfo {volume} {2}},\ \bibinfo {pages} {467} (\bibinfo {year}
  {1985})}\BibitemShut {NoStop}%
\bibitem [{\citenamefont {Yurke}\ and\ \citenamefont
  {Potasek}(1987)}]{Yurke1987}%
  \BibitemOpen
  \bibfield  {author} {\bibinfo {author} {\bibfnamefont {B.}~\bibnamefont
  {Yurke}}\ and\ \bibinfo {author} {\bibfnamefont {M.}~\bibnamefont
  {Potasek}},\ }\href {\doibase 10.1103/PhysRevA.36.3464} {\bibfield  {journal}
  {\bibinfo  {journal} {Physical Review A}\ }\textbf {\bibinfo {volume} {36}},\
  \bibinfo {pages} {3464} (\bibinfo {year} {1987})}\BibitemShut {NoStop}%
\bibitem [{\citenamefont {Tapster}\ and\ \citenamefont
  {Rarity}(1998)}]{Tapster1998}%
  \BibitemOpen
  \bibfield  {author} {\bibinfo {author} {\bibfnamefont {P.~R.}\ \bibnamefont
  {Tapster}}\ and\ \bibinfo {author} {\bibfnamefont {J.~G.}\ \bibnamefont
  {Rarity}},\ }\href {\doibase 10.1080/095003498151852} {\bibfield  {journal}
  {\bibinfo  {journal} {Journal of Modern Optics}\ }\textbf {\bibinfo {volume}
  {45}},\ \bibinfo {pages} {595} (\bibinfo {year} {1998})}\BibitemShut
  {NoStop}%
\bibitem [{\citenamefont {Blauensteiner}\ \emph {et~al.}(2009)\citenamefont
  {Blauensteiner}, \citenamefont {Herbauts}, \citenamefont {Bettelli},
  \citenamefont {Poppe},\ and\ \citenamefont
  {H{\"{u}}bel}}]{Blauensteiner2009}%
  \BibitemOpen
  \bibfield  {author} {\bibinfo {author} {\bibfnamefont {B.}~\bibnamefont
  {Blauensteiner}}, \bibinfo {author} {\bibfnamefont {I.}~\bibnamefont
  {Herbauts}}, \bibinfo {author} {\bibfnamefont {S.}~\bibnamefont {Bettelli}},
  \bibinfo {author} {\bibfnamefont {A.}~\bibnamefont {Poppe}}, \ and\ \bibinfo
  {author} {\bibfnamefont {H.}~\bibnamefont {H{\"{u}}bel}},\ }\href {\doibase
  10.1103/PhysRevA.79.063846} {\bibfield  {journal} {\bibinfo  {journal}
  {Physical Review A - Atomic, Molecular, and Optical Physics}\ }\textbf
  {\bibinfo {volume} {79}},\ \bibinfo {pages} {1} (\bibinfo {year} {2009})},\
  \Eprint {http://arxiv.org/abs/0810.4785} {arXiv:0810.4785} \BibitemShut
  {NoStop}%
\bibitem [{\citenamefont {Hum}\ and\ \citenamefont {Fejer}(2007)}]{Hum2007}%
  \BibitemOpen
  \bibfield  {author} {\bibinfo {author} {\bibfnamefont {D.~S.}\ \bibnamefont
  {Hum}}\ and\ \bibinfo {author} {\bibfnamefont {M.~M.}\ \bibnamefont
  {Fejer}},\ }\href {\doibase 10.1016/j.crhy.2006.10.022} {\bibfield  {journal}
  {\bibinfo  {journal} {Comptes Rendus Physique}\ }\textbf {\bibinfo {volume}
  {8}},\ \bibinfo {pages} {180} (\bibinfo {year} {2007})}\BibitemShut {NoStop}%
\bibitem [{\citenamefont {Dosseva}\ \emph {et~al.}(2016)\citenamefont
  {Dosseva}, \citenamefont {Cincio},\ and\ \citenamefont
  {Bra{\'{n}}czyk}}]{Dosseva2014}%
  \BibitemOpen
  \bibfield  {author} {\bibinfo {author} {\bibfnamefont {A.}~\bibnamefont
  {Dosseva}}, \bibinfo {author} {\bibfnamefont {{\L}.}~\bibnamefont {Cincio}},
  \ and\ \bibinfo {author} {\bibfnamefont {A.~M.}\ \bibnamefont
  {Bra{\'{n}}czyk}},\ }\href {\doibase 10.1103/PhysRevA.93.013801} {\bibfield
  {journal} {\bibinfo  {journal} {Physical Review A}\ }\textbf {\bibinfo
  {volume} {93}},\ \bibinfo {pages} {013801} (\bibinfo {year} {2016})},\
  \Eprint {http://arxiv.org/abs/1410.7714} {arXiv:1410.7714} \BibitemShut
  {NoStop}%
\bibitem [{\citenamefont {Tambasco}\ \emph {et~al.}(2016)\citenamefont
  {Tambasco}, \citenamefont {Boes}, \citenamefont {Helt}, \citenamefont
  {Steel},\ and\ \citenamefont {Mitchell}}]{Tambasco2016}%
  \BibitemOpen
  \bibfield  {author} {\bibinfo {author} {\bibfnamefont {J.-L.}\ \bibnamefont
  {Tambasco}}, \bibinfo {author} {\bibfnamefont {A.}~\bibnamefont {Boes}},
  \bibinfo {author} {\bibfnamefont {L.~G.}\ \bibnamefont {Helt}}, \bibinfo
  {author} {\bibfnamefont {M.~J.}\ \bibnamefont {Steel}}, \ and\ \bibinfo
  {author} {\bibfnamefont {A.}~\bibnamefont {Mitchell}},\ }\href {\doibase
  10.1364/oe.24.019616} {\bibfield  {journal} {\bibinfo  {journal} {Optics
  Express}\ }\textbf {\bibinfo {volume} {24}},\ \bibinfo {pages} {19616}
  (\bibinfo {year} {2016})}\BibitemShut {NoStop}%
\bibitem [{\citenamefont {Christ}\ \emph {et~al.}(2011)\citenamefont {Christ},
  \citenamefont {Laiho}, \citenamefont {Eckstein}, \citenamefont {Cassemiro},\
  and\ \citenamefont {Silberhorn}}]{Christ2011}%
  \BibitemOpen
  \bibfield  {author} {\bibinfo {author} {\bibfnamefont {A.}~\bibnamefont
  {Christ}}, \bibinfo {author} {\bibfnamefont {K.}~\bibnamefont {Laiho}},
  \bibinfo {author} {\bibfnamefont {A.}~\bibnamefont {Eckstein}}, \bibinfo
  {author} {\bibfnamefont {K.~N.}\ \bibnamefont {Cassemiro}}, \ and\ \bibinfo
  {author} {\bibfnamefont {C.}~\bibnamefont {Silberhorn}},\ }\href {\doibase
  10.1088/1367-2630/13/3/033027} {\bibfield  {journal} {\bibinfo  {journal}
  {New Journal of Physics}\ }\textbf {\bibinfo {volume} {13}} (\bibinfo {year}
  {2011}),\ 10.1088/1367-2630/13/3/033027},\ \Eprint
  {http://arxiv.org/abs/1012.0262} {arXiv:1012.0262} \BibitemShut {NoStop}%
\bibitem [{\citenamefont {Shalm}()}]{Shalm}%
  \BibitemOpen
  \bibfield  {author} {\bibinfo {author} {\bibfnamefont {K.}~\bibnamefont
  {Shalm}},\ }\href {www.spdcalc.org} {\enquote {\bibinfo {title}
  {www.spdcalc.org},}\ }\BibitemShut {NoStop}%
\bibitem [{\citenamefont {Audenaert}\ \emph {et~al.}(2007)\citenamefont
  {Audenaert}, \citenamefont {Calsamiglia}, \citenamefont {Munoz-Tapia},
  \citenamefont {Bagan}, \citenamefont {Masanes}, \citenamefont {Acin},
  \citenamefont {Verstraete}, \citenamefont {Mu{\~{n}}oz-Tapia}, \citenamefont
  {Bagan}, \citenamefont {Masanes}, \citenamefont {Acin},\ and\ \citenamefont
  {Verstraete}}]{Audenaert2007}%
  \BibitemOpen
  \bibfield  {author} {\bibinfo {author} {\bibfnamefont {K.~M.~R.}\
  \bibnamefont {Audenaert}}, \bibinfo {author} {\bibfnamefont {J.}~\bibnamefont
  {Calsamiglia}}, \bibinfo {author} {\bibfnamefont {R.}~\bibnamefont
  {Munoz-Tapia}}, \bibinfo {author} {\bibfnamefont {E.}~\bibnamefont {Bagan}},
  \bibinfo {author} {\bibfnamefont {L.}~\bibnamefont {Masanes}}, \bibinfo
  {author} {\bibfnamefont {A.}~\bibnamefont {Acin}}, \bibinfo {author}
  {\bibfnamefont {F.}~\bibnamefont {Verstraete}}, \bibinfo {author}
  {\bibfnamefont {R.}~\bibnamefont {Mu{\~{n}}oz-Tapia}}, \bibinfo {author}
  {\bibfnamefont {E.}~\bibnamefont {Bagan}}, \bibinfo {author} {\bibfnamefont
  {L.}~\bibnamefont {Masanes}}, \bibinfo {author} {\bibfnamefont
  {A.}~\bibnamefont {Acin}}, \ and\ \bibinfo {author} {\bibfnamefont
  {F.}~\bibnamefont {Verstraete}},\ }\href {\doibase
  10.1103/PhysRevLett.98.160501} {\bibfield  {journal} {\bibinfo  {journal}
  {Physical Review Letters}\ }\textbf {\bibinfo {volume} {98}},\ \bibinfo
  {pages} {1} (\bibinfo {year} {2007})},\ \Eprint
  {http://arxiv.org/abs/0610027} {arXiv:0610027 [quant-ph]} \BibitemShut
  {NoStop}%
\bibitem [{\citenamefont {Nock}\ \emph {et~al.}(2011)\citenamefont {Nock},
  \citenamefont {Dahnoun},\ and\ \citenamefont {Rarity}}]{Nock2011}%
  \BibitemOpen
  \bibfield  {author} {\bibinfo {author} {\bibfnamefont {R.}~\bibnamefont
  {Nock}}, \bibinfo {author} {\bibfnamefont {N.}~\bibnamefont {Dahnoun}}, \
  and\ \bibinfo {author} {\bibfnamefont {J.}~\bibnamefont {Rarity}},\ }in\
  \href {\doibase 10.1109/IMTC.2011.5944324} {\emph {\bibinfo {booktitle}
  {Conference Record - IEEE Instrumentation and Measurement Technology
  Conference}}}\ (\bibinfo {year} {2011})\ pp.\ \bibinfo {pages}
  {475--479}\BibitemShut {NoStop}%
\bibitem [{\citenamefont {Levenberg}(1944)}]{Levenberg1944}%
  \BibitemOpen
  \bibfield  {author} {\bibinfo {author} {\bibfnamefont {K.}~\bibnamefont
  {Levenberg}},\ }\href {\doibase 10.1090/qam/10666} {\bibfield  {journal}
  {\bibinfo  {journal} {Quarterly of Applied Mathematics}\ }\textbf {\bibinfo
  {volume} {2}},\ \bibinfo {pages} {164} (\bibinfo {year} {1944})},\ \Eprint
  {http://arxiv.org/abs/arXiv:1011.1669v3} {arXiv:arXiv:1011.1669v3}
  \BibitemShut {NoStop}%
\bibitem [{\citenamefont {Marquardt}(1963)}]{Marquardt1963}%
  \BibitemOpen
  \bibfield  {author} {\bibinfo {author} {\bibfnamefont {D.~W.}\ \bibnamefont
  {Marquardt}},\ }\href@noop {} {\bibfield  {journal} {\bibinfo  {journal}
  {Journal of the Society for Industral and Applied Mathematics}\ }\textbf
  {\bibinfo {volume} {11}},\ \bibinfo {pages} {431} (\bibinfo {year}
  {1963})}\BibitemShut {NoStop}%
\bibitem [{\citenamefont {Gerry}\ and\ \citenamefont
  {Knight}(2004)}]{Gerry2004}%
  \BibitemOpen
  \bibfield  {author} {\bibinfo {author} {\bibfnamefont {C.~C.}\ \bibnamefont
  {Gerry}}\ and\ \bibinfo {author} {\bibfnamefont {P.}~\bibnamefont {Knight}},\
  }in\ \href@noop {} {\emph {\bibinfo {booktitle} {Introductory Quantum
  Optics}}}\ (\bibinfo  {publisher} {Cambridge University Press},\ \bibinfo
  {address} {Cambridge},\ \bibinfo {year} {2004})\ \bibinfo {edition} {1st}\
  ed.,\ Chap.~\bibinfo {chapter} {7}, p.\ \bibinfo {pages} {182}\BibitemShut
  {NoStop}%
\end{thebibliography}%

\clearpage
\appendix

\section{The Covertness Argument}
\label{sec:covertness-argument}

Our protocol aims to replace one single spatial mode of background radiation seen by a target with
light from a broadband SPDC source, i.e. light produced in a spectral superposition of two-mode
squeezers.
To guarantee perfect covertness, the probe beam needs to be indistinguishable from background light
in any degree of freedom.

Here we demonstrate that the photon statistics of one half of a two-mode squeezed state displays an
identical thermal distribution to background light with the same mean photon number.
Subsequently, we discuss the influence of an imperfect spectral overlap with the background light on the
minimum error probability to detect the probe beam among spatial background modes.

\subsection{Covertness in Photon Number Distribution}
\label{sec:covertn-phot-numb}

The Hamiltonian describing the interaction between a strong laser pump field and a non-linear
crystal creating photon pairs can be written as
\begin{multline}
  \label{apep:1}
  \hat H_{SPDC}\, \propto \, \iint d\omega_s \, d \omega_i \, f(\omega_s, \omega_i) \, \hat
  a_s^\dag(\omega_s) \, \hat a_i^\dag(\omega_i) \, + \\ \, \text{h.c.},
\end{multline}
where $f(\omega_s, \omega_i)$ is the complex joint spectral amplitude of signal and idler photons
and $\hat a_j^\dag(\omega_j)$ with $j \in \{s,i\}$ is the creation operator for signal and idler modes
at spectral mode $\omega_i$.
Once we Schmidt decompose $f(\omega_s, \omega_i)$  such that
\begin{equation}
  \label{apep:4}
  f(\omega_s, \omega_i) \, = \, \sum_k \, r_k \, \psi_k(\omega_s) \, \phi_k(\omega_i)
\end{equation}
we can introduce new creation operators for the Schmidt modes and rewrite
\begin{equation}
  \label{apep:3}
  H_{SPDC} \, \propto \, \sum_k \, r_k \, \hat A_{s,k} \, \hat A_{i,k} \, + \, \text{h.c.},
\end{equation}
with
\begin{eqnarray}
  \label{apep:5}
  \hat A_{s,k} & = & \int \, d\omega_s \, \psi_k(\omega_s) \hat a_s(\omega_s)\\
  \hat A_{i,k} & = & \int \, d\omega_i \, \phi_k(\omega_i) \hat a_i(\omega_i).
\end{eqnarray}
See for example \cite{Christ2011}.

We can now identify the solution of the Schrödinger equation for each Schmidt mode in the
Hamiltoninan (\ref{apep:3}) as a two-mode squeezer $\hat S_2$ with
\begin{equation}
  \label{apep:6}
  \hat S_2 ( \xi ) \, = \, \exp \left ( \xi^* \, \hat A_s \, \hat A_i \, + \, \xi \, \hat A_s^\dag \,
    \hat A_i^\dag  \right )
\end{equation}
and complex squeezing parameter $\xi = r \, e^{i\, \theta}$ where we are dropping the index $k$ for
brevity.

Now following Gerry and Knight \cite{Gerry2004} we can calculate the two-mode squeezed state in Fock
basis by solving the eigenvalue problem
\begin{equation}
  \label{apep:20}
  \hat A_s \ket{0,0} \, = \, 0
\end{equation}
which can be rewritten as
\begin{equation}
  \label{apep:22}
  \hat S_2( \xi ) \, \hat A_s \, \hat S_2^\dag( \xi ) \, \hat S_2( \xi ) \, \ket{0,0} \, = \, 0,
\end{equation}
where we multiply $\hat S_2( \xi )$ from the left and use $\hat S_2^\dag( \xi )\, \hat S_2( \xi ) =
\mathds{1}$.
Using a result from the Baker-Hausdorf lemma we can write
\begin{equation}
  \label{apep:23}
  \hat S_2( \xi ) \, \hat A_s \, \hat S_2^\dag( \xi ) \, = \, \hat A_s \, \cosh r \, + \, \
  e^{i\,\theta} \, \hat A_i^\dag \, \sinh r
\end{equation}
and with $\mu = \cosh r$ and $\nu = e^{i \theta} \sinh r$ the eigenvalue problem becomes
\begin{equation}
  \label{apep:24}
  \left ( \mu \, \hat A_s \, + \, \nu \, \hat A_i^\dag \right ) \ket{\xi} \, = \, 0,
\end{equation}
where $\ket{\xi} = \hat S_2(\xi)\, \ket{0,0}$ denotes the two-mode squeezed vacuum state as it is
produced by SPDC.

Choosing the ansatz for $\ket{\xi}$ in Fock representation as
\begin{equation}
  \label{apep:25}
  \ket{\xi} \, = \, \sum_{n,m} \, C_{n,m} \, \ket{n,m}
\end{equation}
and inserting into (\ref{apep:24}) we find
\begin{multline}
  \label{apep:26}
  \sum_{n,m} \, C_{n,m} \left [ \mu \, \sqrt{n} \, \ket{n-1,m} \, + \, \nu \, \sqrt{m+1}\,
    \ket{n,m+1} \right ]\, \\ = \, 0.
\end{multline}
with solution
\begin{multline}
  \label{apep:27}
  C_{n.m} \, = \, C_{0,0} \, \left ( - \frac{\nu}{\mu} \right )^n \, \delta_{n,m} \,\\ = \, C_{0,0} \,
  \left ( -1 \right )^n \, e^{i \, n \, \theta} \, \left ( \tanh r \right )^n \ket{n,n}.
\end{multline}
From normalisation we get $C_{0,0} = \left (\cosh r \right ) ^{-1}$ and can write the two mode
squeezed state as
\begin{equation}
  \label{apep:28}
  \ket \xi \, = \, \frac{1}{\cosh r} \, \sum_{n} \, \left (-1 \right )^n \, e^{i\, n \, \theta} \left (
    \tanh r \right ) ^n \, \ket{n,n}.
\end{equation}
From this we can immediately find the mean photon number in our signal mode $\langle \hat n _s
\rangle =  \langle \hat A_s^\dag \, \hat A_s \rangle =  \sinh ^2 r$.
Furthermore we can calculate the reduced density matrix of our signal mode which is the only mode
available to any object illuminated with our SPDC source
\begin{multline}
  \label{apep:29}
  \hat \rho_s \, = \, \Trace_i\left [ \ket{\xi} \bra{\xi}  \right ]\, \\= \, \frac{1}{\cosh^2 r} \,
  \sum_n \, \left (\tanh r \right ) ^{2n} \ket{n} \bra{n}
\end{multline}

Finally calculating the photon number distribution in our signal mode we find
\begin{equation}
  \label{apep:30}
  P_n^{(s)} \, = \, \bra{n} \rho_s \ket{n} \, = \, \frac{\langle \hat n_s \rangle ^n}{\left ( 1 \, + \,
    \langle \hat n_s \rangle \right ) ^{n+1}}\, ,
\end{equation}
which corresponds to a thermal distribution with mean photon number $\langle \hat n_s \rangle$.

The fact that one single mode of a two mode squeezed state shows this behaviour guarantees the
photon number distribution is correctly matched against other thermal sources which includes all
typical sources of background light such as the sun.
In contrast, it is easy to see that a laser, exhibiting a Poisson distribution in photon number,
conventionally used to illuminate targets in LIDAR systems can never be covert, because a Poissonian
photon number distribution will always differ from a thermal distribution even if the mean photon
number is matched.

\subsection{Spectral Covertness}
\label{sec:spectral-covertness}

Besides the photon statistics, one could also envisage exploiting a mismatched spectral density to
reveal the probe beam of our rangefinder.
Of course if the rangefinder emits light that does not perfectly match the surroundings spectrally,
there will always be a chance for the target to detect this.
While in principle we can use spectral engineering of quasi-phase-matched crystals to perfectly
emulate the same spectrum as the background around the rangefinder, it is challenging to produce and
collect the broadband photons needed to emulate true thermal background.

Yet, we can quantify the probability for the target to detect the rangefinder after $N$ photons
as a function of the spectral overlap between rangefinder and background.
The minimum error probability of discerning $N$ copies of a background state $\hat \rho_b$ from $N$
copies of the probe state $\hat \rho_s$ is given by
\begin{equation}
  \label{apep:2}
  P_e(N) \, = \, \frac{1}{2} \left ( 1- \frac{1}{2} || \hat \rho_b^{\otimes N} - \hat \rho_s^{\otimes N} ||_{\Trace}\right ),
\end{equation}
where $||\hat \rho||_{\Trace}=\Trace\left ( \sqrt{\hat \rho^* \hat \rho} \right )$ is the trace
norm~\cite{Audenaert2007}.
The trace norm is closely related to the fidelity $F$ through
\begin{multline}
  \label{apep:32}
  F(\hat \rho_0, \hat \rho_1 ) \, = \, ||\sqrt{\hat\rho_0}\sqrt{\hat\rho_1}||_{\Trace} \, \\= \,
  \Trace\left ( \sqrt{\left ( \sqrt{\hat\rho_0}\sqrt{\hat\rho_1} \right ) ^* \sqrt{\hat\rho_0}\sqrt{\hat\rho_1}}\right )
\end{multline}
and maintains the inequality
\begin{equation}
  \label{apep:33}
  1-F(\hat \rho_0 , \hat \rho_1 ) \, \leq \, \frac{1}{2} || \hat \rho_0 - \hat \rho_1 ||_{\Trace} \leq \sqrt{1-F(\hat \rho_0 , \hat \rho_1 )^2}.
\end{equation}

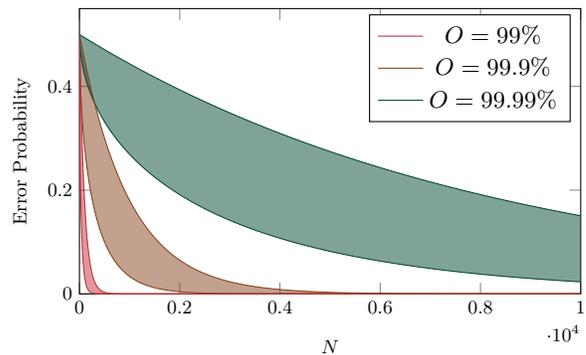
\begin{figure}
  \centering
  \begin{tikzpicture}
    \begin{axis}[
      xmin=0, xmax=10000,
      ymin=0,
      width=.46\textwidth, height=.3\textwidth,
      xlabel={$N$},
      ylabel={Error Probability},
      x label style={at={(axis description cs:0.5,0.0)},anchor=north, scale=0.8},
      y label style={at={(axis description cs:0.1,0.5)},anchor=south, scale=0.8},
      tick label style={scale=0.8}
      ]
      \addplot[name path=top1, uobRed, smooth, samples=800, domain=0:10000]
      {1/2*(1-sqrt(1-.99^(2*x)))};
      \addplot[name path=top2, uobDarkYellow, smooth, samples=800, domain=0:10000]
      {1/2*(1-sqrt(1-.999^(2*x)))};
      \addplot[name path=top3, uobDarkGreen, smooth, samples=800, domain=0:10000]
      {1/2*(1-sqrt(1-.9999^(2*x)))};
      \addplot[name path=bottom1, uobRed, smooth, samples=800, domain=0:10000]
      {1/2*.99^x};
      \addplot[name path=bottom2, uobDarkYellow, smooth, samples=800, domain=0:10000]
      {1/2*.999^x};
      \addplot[name path=bottom3, uobDarkGreen, smooth, samples=800, domain=0:10000]
      {1/2*.9999^x};
      \addplot[uobRed, opacity=.5] fill between[of=bottom1 and top1];
      \addlegendentry{$O=99$\%}
      \addplot[uobDarkYellow, opacity=.5] fill between[of=bottom2 and top2];
      \addlegendentry{$O=99.9$\%}
      \addplot[uobDarkGreen, opacity=.5] fill between[of=bottom3 and top3];
      \addlegendentry{$O=99.99$\%}
    \end{axis}
  \end{tikzpicture}
  \caption{Bounds on the minimal error probability for distinguishing between background and
    rangefinder for spectral overlap $O$. The error probability decreases quickly with the number of
    photons sent towards the target $N$, raising the need for well engineered spectral properties of
    the down-conversion source.}
  \label{fig:min-err-prob}
\end{figure}

Because $\hat \rho_s$ and $\hat \rho_b$ are mixed states with spectral density $f(\omega)$ and
$g(\omega)$, respectively, we can write them as
\begin{eqnarray}
  \label{apep:35}
  \rho_s & = & \int d\omega \, f(\omega) \ket{\omega}\bra{\omega} \\
  \rho_b & = & \int d\omega \, g(\omega) \ket{\omega}\bra{\omega}.
\end{eqnarray}
Furthermore, the fidelity between the two states becomes
\begin{equation}
  \label{apep:36}
  F(\hat \rho_s, \hat \rho_b ) \, = \, \int d\omega \, \sqrt{f(\omega) \, g(\omega)} \, =: \, O,
\end{equation}
with spectral overlap $O$ and the fidelity between $N$ copies of those states yields
\begin{equation}
  \label{apep:37}
  F(\hat \rho_s^{\otimes N}, \hat \rho_b^{\otimes N} ) \, = \, O^N.
\end{equation}
Substituting equation (\ref{apep:37}) into equation (\ref{apep:2}) and using (\ref{apep:33}) we find the
following inequality for the minimum error probability
\begin{equation}
  \label{apep:38}
  \frac{1}{2} \, \left ( 1 - \sqrt{1-O^{2N}} \right ) \leq P_e(N) \leq \frac{1}{2} O^N.
\end{equation}
This gives an upper and lower bound for the minimum probability that the target is wrongly
discerning the rangefinder from background.
We can plot this probability against the number of photons sent towards the target for different
spectral overlaps $O$ (figure \ref{fig:min-err-prob}).
If no photons have been sent out ($N=0$) the best probability the target has is $P_e(0)=0.5$, i.e.
it can only guess whether the spatial mode it is investigating is background light or our
rangefinder.
However its chances improve rapidly with the number of incident photons, with high probability to
distinguish the two states even at $N=5000$ and a spectral overlap of $99.9\%$.

However, saturating this ideal mathematical bound would require the target to have access to a full
quantum optics lab, including quantum memory, in order to perform a collective measurement on all
photons received from the mode it is investigating.
Realistic targets without ideal measurement resources will perform considerably worse, so that the
results shown here should be regarded as the provably secure limits within which covertness is
guaranteed by the laws of nature.
The discrepancy between the potential information gained in a theoretically optimal quantum
measurement and the capabilities of an adversary with presently feasible detection capabilities
still allows significant scope for practical implementations of covert rangefinding, which are
secure within the limits of current technology.
Additionally, while spectrally engineering down-conversion to match the background might be
difficult in certain situations, it is also possible to mix in spectral modes from the true
background (not occupied by the down-conversion source), to again achieve a perfect spectral overlap
between rangefinder and background and thereby perfect and provable covertness.

\section{Signal and Noise Terms}
\label{sec:signal-noise-terms}

In equation~(\ref{apep:3}) we define our SNR model and summarise all occurring noise terms in
different categories in table~\ref{tab:SNRTerms}.
In this section we give the exact formulas for every single noise term.

\subsection{Signal}
\label{sec:signal}

The signal term $S$ in our model consists of photon pairs detected in one time bin after integration
time $T$.
\begin{equation}
  \label{apep:7}
  S \, = \, c_p \, Q \cdot T,
\end{equation}
with photon pair rate $c_p$ and optical loss (gain $Q<1$).
Meaning that photons sent towards the target get lost with probability $Q$.
This reduces the detected coincidence rate to $c_P \, Q$ leaving $S$ photon pair events after
integration time $T$.

\subsection{Proportional Terms}
\label{sec:proportional-terms}

Table~\ref{tab:SNRTerms} lists only one term that is proportional to the number of
detector pairs/frequency bins used in our protocol.
Accidental coincidences caused by the detection of dark counts within a bin width $\Delta t$ are the
only contributors to this noise term.
Hence,
\begin{equation}
  \label{apep:8}
  N_{dd} \, = \, c_d^2 \cdot \Delta t \, T,
\end{equation}
where $c_d$ is the dark count rate of the detectors.
Note how the term is dependant on the bin width $\Delta t$ since, like all other noise terms, dark
counts are a random process.
Consequently they have a finite probability to coincide within this bin width.
Photon pairs in the signal term $S$ are not dependant on this bin width (as long as the detector
jitter is low enough), since they will always coincide.
Accidental coincidences between dark counts are proportional to the number of detectors since every
detector introduces the same dark count rate to the system.
\subsection{Constant Terms}
\label{sec:constant-terms}

We identified six terms in total that contribute a constant amount of noise to our system,
independently of the frequency bin number $n$.
These terms are caused by event combinations where one detector click is caused by a dark count
event.
For example, the combination of a single photon of a detected photon pair in the idler mode and a
dark count in the signal mode occurs
\begin{equation}
  \label{apep:9} \nonumber
  c_d \cdot \frac{c_p}{n} \cdot  \Delta t \, T,
\end{equation}
in a single frequency bin and
\begin{equation}
  \label{apep:10}
  N_{d,c} \, = c_d \cdot Q \, c_p \cdot  \Delta t \, T
\end{equation}
in $n$ frequency bins.

No physical implementation of a photon pair source will ever reach a unity heralding efficiency
$\eta < 1$.
Hence, every photon pair source will also produce photons who's partner was lost between the photon
creation and the detection.
These terms are then, similar like background light from the surrounding, contributing to the noise
as
\begin{equation}
  \label{apep:11}
  N_{d,s} = c_d \cdot Q \, c_s \cdot \Delta t \, T.
\end{equation}
with unpaired photon rate $c_s$ defined at the pair source by the heralding efficiency $\eta = \frac{c_p}{c_p+c_s}$.

Lastly combinations between background light from the environment an dark counts can occur
\begin{equation}
  \label{apep:12}
  N_{d,B} \, = \, c_d \cdot B_0 \, \Delta \lambda \cdot \Delta t \, T
\end{equation}
times, where $B_0$ is the backgrounds spectral density in $\si{\hertz \per \nano \metre}$ and
$\Delta \lambda$ denotes the combined spectral bandwidth of all frequency bins.

Of course these terms also exist for the case where a dark count event was registers in the idler
mode.
However, the magnitudes of the different terms vary greatly.
This is mostly so because it is much less likely to detect a photon from the source (either paired
or unpaired) after it has travelled to the target.
On the other hand background light is typically much higher on the target facing detectors.

\subsection{Inversely Proportional Terms}
\label{sec:invers-prop-terms}

Quantum rangefinding can reduce the typically high background rates associated with broadband single
photon detection by only correlating frequency bins that are constraint under energy conservation.
Events between detectors violating this condition can not have been produced in the down-conversion
process and can hence be omitted.
While this mechanic greatly helps with removing environmental background it is also beneficial
towards other events emerging from imperfect implementations.

The terms most relevant to the advantage gained by omitting non energy constraint
correlations include the typically high environmental background in the idler mode.
\begin{equation}
  \label{apep:13}
  N_{c,B} \, = \, c_p \cdot B_0 \, \Delta \lambda \cdot \Delta t \, T
\end{equation}
denotes contributions from photon pairs in the signal mode and background light in its partner mode.
\begin{equation}
  \label{apep:14}
  N_{s,B} \, = \, c_s \cdot B_0 \, \Delta \lambda \cdot \Delta t \, T
\end{equation}
and
\begin{equation}
  \label{apep:15}
  N_{B,B} \, \approx \, 0
\end{equation}
account for noise consisting of combinations between environmental background and unpaired photons
and background in the signal mode, respectively.
Where, again $N_{B,B}$ can be neglected since background from the down-conversion source is
typically low.
For the same reason
\begin{equation}
  \label{apep:16}
  N_{B,s} \, = \, N_{B,c}\, \approx \, 0.
\end{equation}

Quantum rangefinding also helps to reduce noise emerging due to a non-unity heralding efficiency
$\eta < 1$ .
\begin{equation}
  \label{apep:17}
  N_{c,s} \, = \, c_p \cdot Q \, c_s \cdot \Delta t \, T,
\end{equation}
\begin{equation}
  \label{apep:18}
  N_{s,c} \, = \, c_s \cdot Q \, c_p \cdot \Delta t \, T,
\end{equation}
and
\begin{equation}
  \label{apep:19}
  N_{s,s} \, = \, Q \, c_s^2 \cdot \Delta t \, T
\end{equation}
account for accidental coincidences between photon pairs and unpaired photons, unpaired photons and
photon pairs as well as unpaired photons in both modes.
The inversely proportional relation of these terms can be easily seen.
For example in $N_{s,s}$, $c_s$ is the rate of unpaired photons in the source.
Consequently, $\frac{c_s}{n}$ unpaired photon rate remains per frequency bin and yields for each
detector pair constraint under energy conservation $Q \, \left (\frac{c_s}{n} \right )^2 \cdot
\Delta t \, T$ events,
introducing $n$ photon pairs leaves
\begin{equation}
  \label{apep:21}
  n \cdot Q \, \left (\frac{c_s}{n} \right )^2 \cdot \Delta t \, T \, = \, \frac{1}{n} \, N_{s,s}.
\end{equation}

\section{Electronics and Temperature Stabilisation}
\label{sec:electr-temp-stab}

To control the pump laser diode temperature and current as well as the temperature of the
down-conversion crystal, we employed custom build electronics integrated in a small form factor of
$(160 \times 220 \times 52) \, \si{\milli\meter\cubed}$.
\begin{figure}
  \centering
  \begin{tikzpicture}
    \begin{axis}[
      xlabel ={Time [$\si{\second}$]},
      ylabel ={Temperature [$\si{\celsius}$]},
      xmin=0, xmax=940,
      ymin=15, ymax=60,
      legend pos = south east
      ]
      \addplot[uobDarkAqua, smooth, forget plot] table [
      x expr = \thisrowno{0}*0.15,
      y expr = \thisrowno{3}/10.
      ]
      {data/12V-OvenTempLog-Rev1Board.log};
      \addplot[uobBrightAqua, domain=0:200] {0.22*x+18.5};
      \addlegendentry{$13.46\,\si{\celsius\per\minute}$}
    \end{axis}
  \end{tikzpicture}
  \caption{Temperature curve of the crystal oven. The setpoint is reached within $5\,\si{\minute}$
    and is stable to $0.1 \, \si{\celsius}$. The heating slope $\approx 13.5\,\si{\celsius\per\minute}$.}
  \label{fig:oven-temp}
\end{figure}
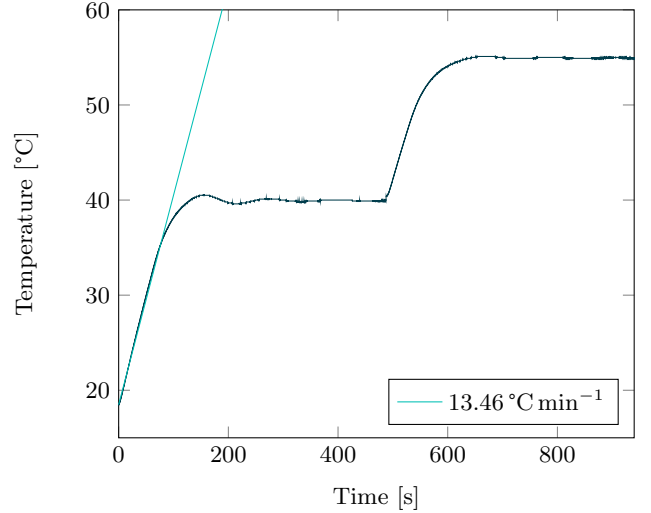
These electronics were capable of stabilising the crystal temperature to $\pm\, 0.1 \,
\si{\celsius}$ as shown in figure~\ref{fig:oven-temp}.
Stabilisation of both, the down-conversion crystal and the laser diode, are necessary to avoid
spectral drifts in the down-converted photons and to guarantee consistent energy correlations
between different frequency bins.

\section{Signal \& Idler Spectra}
\label{sec:signal--idler}
To verify the broadband spectral properties of the down-converted signal and idler beams, they were
measured independently from each other on a single photon resolving spectrometer.
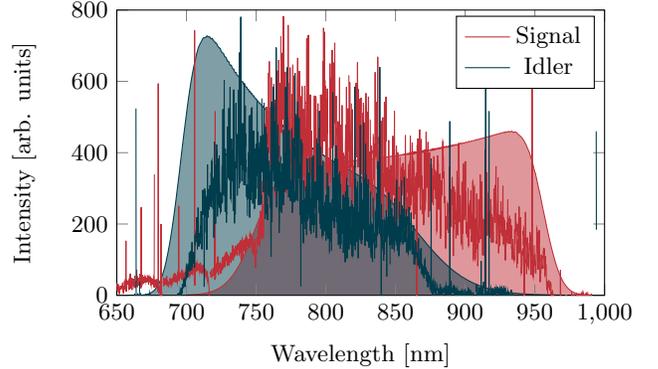
\begin{figure}
  \centering
  \begin{tikzpicture}
    \begin{axis}[
      width=0.45\textwidth, height=0.3\textwidth, hide x axis, hide y axis, xmin = 650, xmax=1000, ymin=0]
      \addplot[uobRed, name path=ssim, smooth] table[col sep=comma, smooth, x index=0, y index=1]
      {data/crystalB-simulation.csv};
      \addplot[uobDarkAqua, name path=isim, smooth] table[col sep=comma, smooth, x index=2, y index=3]
      {data/crystalB-simulation.csv};
      \path[name path=axis] (axis cs:650,0) -- (axis cs:1000,0);
      \addplot [uobRed, opacity=0.5] fill between [of=ssim and axis];
      \addplot [uobDarkAqua, opacity=0.5] fill between [of=isim and axis];
    \end{axis}
    \begin{axis}[
      width=0.45\textwidth, height=0.3\textwidth,
      xmin=650, xmax=1000,
      restrict x to domain=650:1000,
      ymin=0, ymax=800,
      restrict y to domain=-20:800,
      xlabel={Wavelength [\si{\nano\metre}]},
      ylabel={Intensity [arb. units]}
      ]
      \addplot[uobRed, smooth] table[col sep=comma,
      x index = 0,
      y expr = \thisrowno{1}-\thisrowno{3}-100
      ]
      {data/crystalB-complete_mod4.csv};
      \addlegendentry{Signal}
      \addplot[uobDarkAqua, smooth] table[col sep=comma,
      x index = 0,
      y expr = \thisrowno{2}-\thisrowno{3}-100
      ]
      {data/crystalB-complete_mod4.csv};
      \addlegendentry{Idler}
    \end{axis}
  \end{tikzpicture}
  \caption{Results of the signal and idler spectra, showing broadband phase-matching.
    The filled plots describe the calculations using our software. The solid lines are the data
    measured with the single-photon spectrometer.}
  \label{fig:crystalB-spectra}
\end{figure}
Both spectra are shown in figure~\ref{fig:crystalB-spectra}.
The solid backdrops show the calculations performed with our software while the lines in the
foreground show the measured spectral density.
Originally both spectra were designed to be identical.
However, a mismatch between the laser diode wavelength and the pump wavelength the phase-matching
was designed for causes a red and blue shift for signal and idler photons, respectively.

\end{document}